\documentclass[journal,twoside,web]{ieeecolor}
\usepackage{tmi}
\usepackage{cite}
\usepackage{amsmath,amssymb,amsfonts,caption}
\usepackage{algorithmic}
\usepackage{graphicx}
\usepackage{textcomp}
\usepackage{booktabs}
\usepackage{color}
\usepackage{hyperref}
\usepackage{multirow}
\usepackage{booktabs}

\usepackage{fancyhdr}

\def \ourmethod {LE-UDA}

\def\BibTeX{{\rm B\kern-.05em{\sc i\kern-.025em b}\kern-.08em
    T\kern-.1667em\lower.7ex\hbox{E}\kern-.125emX}}
\begin{document}

\title{LE-UDA: Label-efficient unsupervised domain adaptation for medical image segmentation}
\author{Ziyuan Zhao,~\IEEEmembership{Member, IEEE}, Fangcheng Zhou, Kaixin Xu, Zeng Zeng,~\IEEEmembership{Senior Member, IEEE}, Cuntai Guan,~\IEEEmembership{Fellow,~IEEE}, S. Kevin Zhou,~\IEEEmembership{Fellow,~IEEE}
\thanks{Corresponding authors: Zeng Zeng, Cuntai Guan.}
\thanks{Z. Zhao is with the School of Computer Science and Engineering, Nanyang Technological University, and also with the Institute for Infocomm Research (I$2$R) \& Artificial Intelligence, Analytics And Informatics (AI$^3$), Agency for Science, Technology and Research (A*STAR), Singapore. E-mail: zhaoz AT i2r.a-star.edu.sg.}
\thanks{F. Zhou is with National University of Singapore, Singapore. This work was done when F. Zhou did internship with I2R, A*STAR. E-mail: fangchengzhou AT outlook.com.}
\thanks{K. Xu is with the Institute for Infocomm Research (I$2$R), Agency for Science, Technology and Research (A*STAR), Singapore. E-mail: xuk AT i2r.a-star.edu.sg.}
\thanks{Z. Zeng is with the School of Microelectronics, Shanghai University, China, and also with the Institute for Infocomm Research (I$2$R), Agency for Science, Technology and Research (A*STAR), Singapore. E-mail: zengz AT shu.edu.cn.}
\thanks{C. Guan is with the School of Computer Science and Engineering, Nanyang Technological University, Singapore. E-mail: ctguan AT ntu.edu.sg.}
\thanks{S. Zhou is with the Center for Medical Imaging, Robotics, Analytic Computing and Learning (MIRACLE), School of Biomedical Engineering \& Suzhou Institute for Advanced Research, University of Science and Technology of China, Suzhou, China, and also with the Key Laboratory of Intelligent Information Processing of Chinese Academy of Sciences (CAS), Institute of Computing Technology, CAS, Beijing, China. E-mail: skevinzhou AT ustc.edu.cn.}
}
\maketitle

\thispagestyle{fancy}
\fancyhead{}
\lhead{}
\lfoot{\scriptsize{© 2022 IEEE. Personal use of this material is permitted. Permission from IEEE must be obtained for all other uses, in any current or future media, including reprinting/republishing this material for advertising or promotional purposes, creating new collective works, for resale or redistribution to servers or lists, or reuse of any copyrighted component of this work in other works.}}
\cfoot{}
\rfoot{}

\begin{abstract}
While deep learning methods hitherto have achieved considerable success in medical image segmentation, they are still hampered by two limitations: (i) reliance on large-scale well-labeled datasets, which are difficult to curate due to the expert-driven and time-consuming nature of pixel-level annotations in clinical practices, and (ii) failure to generalize from one domain to another, especially when the target domain is a different modality with severe domain shifts.
Recent unsupervised domain adaptation~(UDA) techniques leverage abundant labeled source data together with unlabeled target data to reduce the domain gap, but these methods degrade significantly with limited source annotations.
In this study, we address this underexplored UDA problem, investigating a challenging but valuable realistic scenario, where the source domain not only exhibits domain shift~w.r.t. the target domain but also suffers from label scarcity. In this regard, we propose a novel and generic framework called ``Label-Efficient Unsupervised Domain Adaptation"~(LE-UDA). In LE-UDA, we construct self-ensembling consistency for knowledge transfer between both domains, as well as a self-ensembling adversarial learning module to achieve better feature alignment for UDA.
To assess the effectiveness of our method, we conduct extensive experiments on two different tasks for cross-modality segmentation between MRI and CT images. Experimental results demonstrate that the proposed LE-UDA can efficiently leverage limited source labels to improve cross-domain segmentation performance, outperforming state-of-the-art UDA approaches in the literature.

\end{abstract}

\begin{IEEEkeywords}
Unsupervised domain adaptation, medical image segmentation, cross-modality learning, semi-supervised learning, adversarial learning
\end{IEEEkeywords}

\section{Introduction}
\begin{figure}[t]
    \centering
    \includegraphics[width=0.9\linewidth]{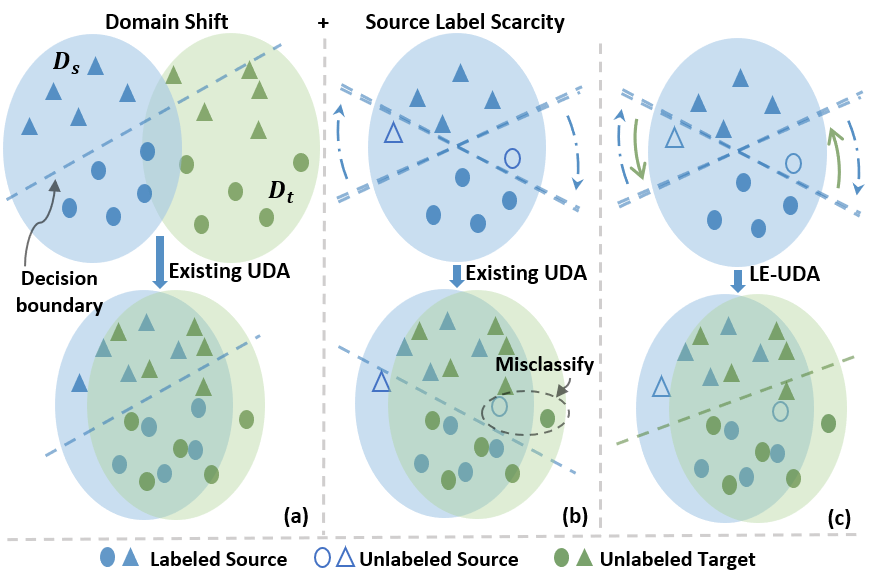}
    \caption{Schematic diagrams of the consequences of source label scarcity when unsupervised domain adaptation. \textbf{Left:} Existing UDA methods align source and target distributions given the abundant source labels. \textbf{Middle:} Under source label scarcity, existing UDA methods may lead to suboptimal solutions with the fuzzy decision boundary. \textbf{Right:} LE-UDA~can correct the fuzzy decision boundary by leveraging unlabeled data, thereby improving cross-domain performance.}
    \label{fig:intro}
\end{figure}

\label{sec:introduction}
\IEEEPARstart{R}{ecent} advancements of deep learning~\cite{zhou2019handbook,zhou2021review} have shown tremendous success in medical image segmentation~\cite{ronneberger2015u, cciccek20163d, milletari2016v}. 
Nevertheless, these impressive achievements always come with the price of massive pixel-accurate annotations, which are both costly and labor-intensive to obtain, thereby leading to the label scarcity problem in clinical practice~\cite{tajbakhsh2020embracing}.
To mitigate this problem, many recent efforts in medical image analysis have been devoted to developing methodologies beyond fully supervised learning techniques, such as self-supervised~\cite{bai2019self, li2020self,zhou2021models,sesenet,ZHU2020101746}, semi-supervised ~\cite{bai2017semi,zhao2019semi,zhao2021dsal}, and weakly-supervised learning~\cite{rajchl2016deepcut,playout2019novel, belharbi2021deep}. 
Nevertheless, most of these not-so-supervised methods are designed for a single-domain scenario and neglect that in real-world clinical scenarios, data can be collected from a variety of sources, including different medical centers, subject cohorts, imaging scanners, protocols, and even modalities, such as MRI and CT, resulting in different data distributions across domains~\cite{gibson2018inter,9557808}.
As a result, the model trained on one domain, \emph{i.e.}, source domain, usually fails in generalizing well to the data collected from unseen domains, \emph{i.e.}, target domain, crippling the model performance significantly.
To improve cross-domain generalization in the existence of domain shift, one simple but effective method is to use newly annotated data from the target domain to fine-tune the model trained on the data from the source domain~\cite{ghafoorian2017transfer}. 
However, labeling data from new domains inevitably involves additional time, expenses, and costs. 

To address the well-known domain shift phenomena in the absence of target annotations, significant progress has been made for unsupervised domain adaptation~(UDA)~\cite{zhu2017unpaired,russo2018source,zhang2018task, bousmalis2017unsupervised, zhao2018supervised, ganin2016domain, tzeng2017adversarial,hoffman2018cycada,chen2019synergistic}.
In a common UDA setting, most previous efforts assume that abundant source annotations are available and these UDA methods aim to minimize domain discrepancy by means of utilizing labeled data from the source domain and unlabeled data from the target domain, thereby improving cross-domain adaptation performance.
However, in clinical practice, such assumptions cannot always hold since the source domain may also suffer from label scarcity due to various reasons,~\emph{e.g.}, limited access and expert knowledge. 
Source label scarcity can be detrimental to existing UDA methods, causing a significant drop in performance. This is further illustrated in Fig.~\ref{fig:intro}. In the absence of sufficient source annotations, the models cannot be well trained on the source domain, resulting in a suboptimal decision boundary, which subsequently influences domain alignment. In the experiment section, we shall further demonstrate this circumstance.

As such, this problem motivates us to study a practical yet challenging UDA scenario, \emph{i.e.}, UDA under source label scarcity, where only scarce source annotations are available. 
It is earlier noted that existing UDA methods cannot handle this problem, as they require full source labels without leveraging unlabeled source data, except the prior work MT-UDA~\cite{zhao2021mt}, which introduced semi-supervised learning~(SSL) into UDA and proposed a self-ensembling-based UDA framework to realize cross-modality medical image segmentation. 
In MT-UDA, generative adversarial networks~(GANs)~\cite{GAN} were used to align cross-domain distributions at the image level and also derive synthetic images for both source and target domains.
Subsequently, self-ensembling models were built to explore the knowledge from real and synthetic images with diversified distributions for improving cross-domain performance under source label scarcity. However, MT-UDA solely enforces structural consistency across domains while ignoring adversarial learning for better feature alignment.
Therefore, MT-UDA was imperfect as the feature adaptation has not been adequately investigated.

In this work, we propose a label-efficient unsupervised domain adaptation framework, named \ourmethod, for tackling domain shift under source label scarcity. We first leverage the strong generative capacity of GANs to facilitate image adaptation and generate complementary domain images to diversify the training distributions, thereby improving the model's generalization ability. 
To explore the rich information behind diverse domains, a dual-teacher network is constructed based on self-ensembling~\cite{meanteacher}, in which the student model exploits intra-domain knowledge from source images and synthetic source images while also distilling inter-domain knowledge from a diversity of complementary domains. 
To further motivate the student model to more comprehensively explore domain-invariant features, we engage adversarial learning into the teacher-student learning framework for feature adaptation. 
To benefit from both teacher-student learning and adversarial learning, we propose a dual self-ensembling adversarial learning mechanism, in which discriminators are connected to multi-level predictions of the student and teacher networks under explicit self-ensembling consistency to promote the student network to derive generalizable representations implicitly.
As a result, our proposed method can effectively transfer multi-domain knowledge from diverse domains, achieving label-efficient unsupervised domain adaptation against source label scarcity.
To sum up, the main contributions of this paper are as follows:

\begin{itemize}
    \item We investigate a new UDA scenario in which only limited source labels are accessible. The scenario is practical and challenging for real-world applications. We present a novel UDA approach, \ourmethod, to address this issue. In~\ourmethod, we jointly exploit (i) both image and feature adaptations to address domain shift, and (ii) both inter-domain and intra-domain knowledge transfer to handle source label scarcity.
    
    \item To effectively leverage the rich information behind different domains, we propose a dual-domain consistency paradigm based on self-ensembling to extensively mine the dual-domain knowledge for label-efficient cross-domain image segmentation. Moreover, a dual adversarial self-ensembling learning strategy is designed to enhance the feature alignment and facilitate reliable dual-domain knowledge transfer.
    
    \item We carry out extensive studies on two challenging tasks,~\emph{i.e.}, cardiac substructure segmentation and abdominal multi-organ segmentation for bidirectional cross-modality medical image segmentation between CT and MRI. The comprehensive results and analysis reveal that the proposed LE-UDA framework demonstrates superior adaptation performance over the existing state-of-the-art UDA approaches.

\end{itemize}

The work substantially extends our previous work MT-UDA~\cite{zhao2021mt} with the following improvements. First, we improve our method by incorporating self-ensembling adversarial feature alignment. Second, we extensively evaluate our framework for bidirectional cross-modality adaptation with different source label ratios. Third, an additional task~\emph{i.e.}, abdominal multi-organ segmentation, is investigated for a more thorough evaluation. Forth, more UDA methods are compared to show the superiority of the proposed LE-UDA. Last but not least, a comprehensive analysis is provided to demonstrate the feasibility of \ourmethod~for the UDA problem.

\section{Related Work}
\label{sec:related}
\subsection{Unsupervised domain adaptation}
Because of various data distributions induced by various reasons, such as different modalities, domain shift has become a fundamental and common issue in medical imaging analysis and applications. 
With that, domain adaptation~(DA) has long been of great interest.
Based on the levels of label availability on the target domain, existing DA methods can be divided into three types,~\emph{i.e.}, supervised domain adaptation (SDA) with abundant target labels, semi-supervised domain adaptation (SSDA) with limited target labels, and unsupervised domain adaptation (UDA) with no target labels~\cite{9557808}. Compared to SDA and SSDA, which both require additional labeled target data, UDA is more challenging, attractive, and practical since no target labels are required. 
In this context, UDA is becoming increasingly important in the medical image segmentation field, and as a result, a myriad of UDA approaches have been developed for cross-domain medical image segmentation with promising adaptation results~\cite{dou2018pnp,pnp_1,huo2018synseg,chen2019synergistic,chen2020unsupervised,lyu20213}.

Recent years have seen tremendous success with generative adversarial networks (GANs) in image translation and domain adaptation~\cite{GAN,zhu2017unpaired}.
Numerous UDA approaches have been developed to narrow domain discrepancy by implicitly driving alignment via various adversarial learning strategies from different perspectives, including image adaptation~\cite{russo2018source,zhang2018task,bousmalis2017unsupervised,chartsias2017adversarial,huo2018synseg}, feature adaptation~\cite{ganin2016domain,tzeng2017adversarial,tsai2018learning,pnp_1,dou2018pnp} and their mixtures~\cite{hoffman2018cycada,chen2019synergistic,chen2020unsupervised}. 
To achieve image adaptation, many UDA methods have been developed based on CycleGAN~\cite{bousmalis2017unsupervised} with different constraints for unpaired image-to-image transformation, and then leverage synthetic images for training or testing. Chartsias~\emph{et al.}~\cite{chartsias2017adversarial} proposed to harness CycleGAN for CT-to-MRI synthesis, and the synthetic MRI images can be used to train an MRI segmentation model. 
Huo~\emph{et al.}~\cite{huo2018synseg} proposed SynSeg-Net for multi-organ segmentation by combining CycleGAN with a segmentation model together to achieve end-to-end adaptation. 
For feature adaptation, contemporary UDA methods basically draw on the concept of the Domain Adversarial Neural Network (DANN)~\cite{ganin2016domain} and attempt to extract domain-invariant representations via adversarial learning to fool the discriminators in the feature space. 
Dou~\emph{et al.}~\cite{pnp_1} proposed an adversarial domain adaptation network, PnP-AdaNet, for cross-modality cardiac segmentation, in which specific layers can be fine-tuned via adversarial learning to align features in a plug and play fashion, while Tsai~\emph{et al.}~\cite{tsai2018learning} employed adversarial learning to align output label distributions cross domains at various feature levels in the semantic prediction space.

Recent evidence suggests that image and feature adaptation are mutually beneficial, and several attempts have been made to investigate the combination of these two adaptive strategies for boosting adaptation performance. 
Hoffman~\emph{et al.}~\cite{hoffman2018cycada} proposed a method called CyCADA, which combines these two strategies sequentially to achieve more robust domain adaptation performance on natural images. 
Furthermore, Chen~\emph{et al.}~\cite{chen2019synergistic,chen2020unsupervised} designed a UDA framework for synergistic image and feature alignment, called SIFA, achieving promising performance for cross-modality medical image segmentation. Finally, disentangled feature learning is also proposed to mitigate the domain shift between CBCT and CT for vertebra segmentation from CBCT~\cite{lyu20213}. However, contemporary UDA methods heavily rely on training with a great number of source annotations, and their performance may be dramatically degraded when only scarce source annotations are provided, limiting their applications in clinical deployment.
To tackle the source label scarcity problem, we are motivated to explore unlabeled source data for improved model performance using the semi-supervised learning paradigm.

\subsection{Semi-supervised learning}
Label scarcity has been a long-running issue in medical image segmentation, which has been extensively studied on improving performance with limited labels~\cite{cheplygina2019not}.
Recently, the semi-supervised learning~(SSL) paradigm has drawn considerable attention because of its impressive success in leveraging unlabeled data to alleviate the annotation scarcity problem.
Lee~\emph{et al.}~\cite{lee2013pseudo} established a self-training pseudo-labeling framework for semi-supervised learning, in which pseudo labels predicted by a pre-trained model are used to update the model for improving SSL performance iteratively.
Several attempts have been made to combine the pseudo-labeling scheme with different proposed methods to further improve the performance~\cite{bai2017semi,zhao2019semi,zhao2021dsal}. 
For instance, Bai~\emph{et al.}~\cite{bai2017semi} proposed to refine pseudo annotations with the application of CRF for cardiac segmentation from MRI images. However, the self-training process heavily relies on the quality of pseudo labels. Moreover, alternating between training and inference frequently is computationally expensive.

More recently, self-ensembling methods have made remarkable success in semi-supervised image segmentation, achieving promising performance on a wide range of SSL benchmarks. 
Laine and Aila~\cite{tepens2017} proposed Temporal Ensembling (TE) to maintain the temporal consistency using exponential moving averages (EMAs) of the estimates in previous epochs as teacher predictions. 
Tarvainen and Valpola~\cite{meanteacher} extended TE and proposed the mean teacher (MT) framework to ensure prediction consistency of inputs with various perturbations between the student model and the EMA teacher model, thus regularizing the training process under the teacher-student learning scheme.
Subsequently, many approaches~\cite{yu2019uncertainty,li2020transformation,su2019local,li2021hierarchical,luo2021semi} proposed different improvements, including Uncertainty-aware Consistency~\cite{yu2019uncertainty}, Transformation Consistency~\cite{li2020transformation}, Structural Consistency~\cite{su2019local}, Multi-level Consistency~\cite{li2021hierarchical} and Multi-task Consistency~\cite{luo2021semi}. Despite the great success of self-ensembling methods in mitigating the label scarcity problem, most studies have only focused on the single-domain semi-supervised learning scenario, leaving rich cross-domain information unexplored. 
Li~\emph{et al.}~\cite{li2020dual,li2020dual_2} introduced teacher-student learning to explore cross-domain knowledge to address the paucity-of-label problem on the target domain for SSDA but did not take account of the source label scarcity problem for UDA. In contrast, we advance teacher-student learning into UDA to address source label scarcity.

\section{Methodology}
\label{sec:methodology}

\begin{figure*}[t]
    \centering
    \includegraphics[width = 0.8\textwidth]{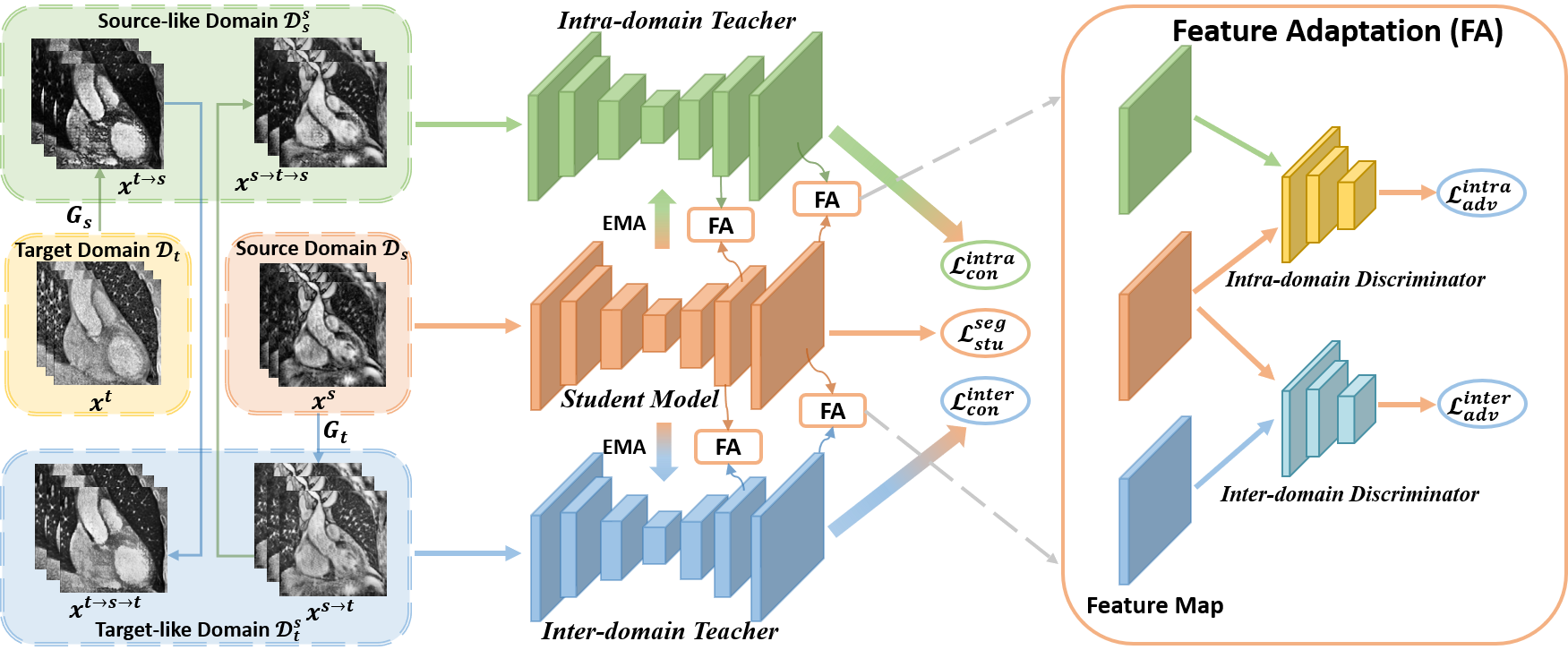}
    \caption{Overview of the proposed \ourmethod~architecture: First, multiple complementary domain images are generated with a dual cycle alignment module (left). Then, these diverse domains are fed into the dual-teacher network for dual-domain knowledge transfer (middle). Simultaneously, inter-domain discriminator and intra-domain discriminator differentiate the extracted features at different levels to derive adversarial losses, which are then back-propagated to the student model for feature adaptation (FA) via self-ensembling adversarial learning (right).}
    \label{fig:archi}
\end{figure*}

In a UDA setting, there are two different domains,~namely source domain $\mathcal{D}_{s}$ and target domain $\mathcal{D}_{t}$.
$\mathcal{D}_{s}$ contains labeled data  $\mathcal{D}_{s}^l=\left\{\left(\mathbf{x}_{i}^{s}, y_{i}^{s}\right)\right\}_{i=1}^{n_l}$ with $n_l$ samples, and unlabeled data $\mathcal{D}_{s}^u=\left\{\left(\mathbf{x}_{i}^{s}\right)\right\}_{i=n_l+1}^{n_l + n_u}$ with $n_u$ samples.
$\mathcal{D}_{t}$ only contains unlabeled data as $\mathcal{D}_{t}=\left\{\left(\mathbf{x}_{i}^{t}\right)\right\}_{i=1}^{m_u}$ with $m_u$ samples. Note that $n_l < n_u$ is assumed in the setting of source label scarcity. The source label ratio $L_{s}$ is denoted as $n_l/(n_l+n_u)$. 
We aim to exploit $\mathcal{D}_{s}^l$, $\mathcal{D}_{s}^u$ and $\mathcal{D}_{t}$ to address the severe domain shift, as well as source label scarcity, thereby improving domain adaptation performance. Fig.~\ref{fig:archi} depicts an overview of the proposed \ourmethod~architecture. In~\ourmethod, two mean-teacher networks are constructed for dual-domain knowledge transfer in an adversarial semi-supervised learning fashion. Our framework includes three major components: (1) generating complementary domain data for image alignment; (2) integrating inter-domain and intra-domain knowledge to assist domain adaptation and supervised training; and (3) enhancing the feature alignment via adversarial teacher-student learning. In the following, we introduce our method in detail from these three perspectives. The notations are summarized in Table~\ref{tab:notations}.

\begin{table}[h]
\centering
\caption{Summary of notations}
\scalebox{0.86}{
\begin{tabular}{ll} 
\toprule
Notation                                       & Definition                               \\ 
\hline
$\mathcal{D}_{s}$, $\mathcal{D}_{t}$ & Image sets of source and target domains  \\
$\mathcal{D}_{s}^l$, $\mathcal{D}_{s}^u$ &  Image sets of labeled and unlabeled source domains       \\
$\mathcal{D}_{s}^s$, $\mathcal{D}_{t}^s$ &  Image sets of source-like and target-like domains           \\
$x^{s}, x^{t}$          & Samples of source and target domains                                         \\
$x^{t \rightarrow s}, x^{s \rightarrow t\rightarrow s}$          & Synthetic source-like and cycle source samples                                         \\
$x^{s \rightarrow t}, x^{t \rightarrow s \rightarrow t}$          & Synthetic target-like and cycle target samples                                         \\
$y^{s}, y^{s \rightarrow t}$          & Source and synthetic target-like labels                                       \\
$G_{s}$, $G_{t}$  & Generators for source-like and target-like images                                \\
$D_{s}$, $D_{t}$  & Discriminators for source and target domains                                  \\
$G$  & The student segmentation network                                 \\
$G_{intra}$, $G_{inter}$  & Intra-domain and inter-domain teachers                                 \\
$D_{intra}$, $D_{inter}$  & Intra-domain and inter-domain discriminators                                 \\
$D^{i}$  & Discriminator for different layers                               \\
\bottomrule
\end{tabular}
}
\label{tab:notations}
\end{table}

\subsection{Dual cycle alignment for image adaptation}

Given that data from different domains, like CT and MRI, exhibit significant image-level distribution heterogeneity, we first adopt a dual cycle alignment module~(DCAM) built on CycleGAN~\cite{zhu2017unpaired} for appearance alignment. In DCAM, two generators of $G_{s}$ and $G_{t}$ are employed to generate synthetic samples by adapting $\mathcal{D}_{s}$ and $\mathcal{D}_{t}$ to each other, while two discriminators of $D_{s}$ and $D_{t}$ are implemented to distinguish whether the images are synthetic or real. More specifically, the target generator $G_{t}$ aims to generate target-like images, \emph{i.e.}, $G_{t}\left(x^{s}\right)=x^{s \rightarrow t}$, while the target discriminator $D_{t}$ seeks to differentiate between the real and fake images of the target domain in a two-player adversarial game and the adversarial loss $\mathcal{L}_{gan}^{t}$ is defined as:
\begin{equation}
\begin{aligned}
\mathcal{L}_{gan}^{t}\left(G_{t}, D_{t}\right)=& \mathbb{E}_{x^{t} \sim \mathcal{D}_{t}}\left[\log D_{t}\left(x^{t}\right)\right]+ \\
&\mathbb{E}_{x^{s} \sim \mathcal{D}_{s}}\left[\log \left(1-D_{t}\left(G_{t}\left(x^{s}\right)\right)\right)\right], 
\end{aligned}
\end{equation}
with which, gradients are propagated from $D_{t}$ to $G_{t}$, encouraging $G_{t}$ to generate images processing similar style to the target domain. Likewise, the pair of $[D_{s},G_{s}]$ is trained with the adversarial loss $\mathcal{L}_{gan}^{s}$, simultaneously. We follow the implementation of CycleGAN~\cite{zhu2017unpaired} and enforce cycle-consistency by introducing an $L_{1}$ penalty on the reconstruction error. The cycle-consistency loss is defined as:
\begin{equation}
\begin{aligned}
\mathcal{L}_{\mathrm{cyc}}\left(G_{s}, G_{t}\right) &=\mathbb{E}_{x_{s} \sim \mathcal{D}_{s}}\left[\left\|G_{s}\left(G_{t}\left(x_{s}\right)\right)-x_{s}\right\|_{1}\right] \\
&+\mathbb{E}_{x_{t} \sim \mathcal{D}_{t}}\left[\left\|G_{t}\left(G_{s}\left(x_{t}\right)\right)-x_{t}\right\|_{1}\right].
\end{aligned}
\end{equation}

With the joint optimization of $\mathcal{L}_{gan}$ and $\mathcal{L}_{\mathrm{cyc}}$, the DCAM can perform unpaired image-to-image translation in a bidirectional fashion, generating synthetic images for appearance alignment. It is possible to divide the synthetic images into two complementary domains,~\emph{i.e.}, source-like domain $\mathcal{D}_{s}^s = \{ x^{t \rightarrow s}, x^{s \rightarrow t \rightarrow s} \}$  and target-like domain and $\mathcal{D}_{t}^s = \{ x^{s \rightarrow t}, x^{t \rightarrow s \rightarrow t} \}$. Except for bridging the image-level domain gap, these newly-augmented intermediate domains retain rich information which can be further explored for SSL and UDA.  For clarity, we skip the illustration of discriminators $D_{s}$, $D_{t}$, and cycle-consistency loss $\mathcal{L}_{\mathrm{cyc}}$ in Fig.~\ref{fig:archi}.

\subsection{Dual-domain knowledge transfer}

\subsubsection{Intra-domain semantic knowledge transfer}
The source-like domain $\mathcal{D}_{s}^s$ includes synthetic source data $x^{t \rightarrow s}$ and cycle source data $x^{s \rightarrow t \rightarrow s}$, with a similar visual appearance like source data $x^{s}$, which allows us to explore the knowledge of $\mathcal{D}_{s}^s$ for improving the within-domain segmentation performance on $\mathcal{D}_{s}$ under source label scarcity. Inspired by the success of mean-teacher~\cite{meanteacher} in utilizing unlabeled data, we construct an intra-domain teacher network $G_{intra}$, which follows the same architecture as the student model $G$ based on self-ensembling~\cite{meanteacher}. 
To be more specific, at training epoch $t$, the intra-domain teacher model weights $f_{\theta^{\prime}}$ are updated using the exponential moving average (EMA) weights of the student model $f_{\theta}$,~\emph{i.e.}, $\theta_{t}^{\prime}=\alpha \theta_{t-1}^{\prime}+(1-\alpha) \theta_{t}$, where $\alpha$ is the EMA decay rate which is used to regulate the impact of student model parameters.
We diversify the inputs of the mean-teacher network by feeding $\mathcal{D}_{s} (\mathcal{D}_s^l, \mathcal{D}_s^u)$ and $\mathcal{D}_{s}^s$ with different perturbations such as noises $\xi$ and $\xi^{\prime}$ to the student and the intra-domain teacher, respectively and encourage their predictions to be consistent by reducing the discrepancy with a mean squared error (MSE) loss $\mathcal{L}_{{con}}^{{intra}}$ as: 
\begin{equation}
\label{eq:loss_kd}
\mathcal{L}_{{con}}^{intra}=\frac{1}{N} \sum_{i=1}^{N}\left\|f\left(x_{i} ; \theta_t^{\prime}, \xi^{\prime}\right)-f\left(x_{i} ; \theta_t, \xi\right)\right\|^{2},
\end{equation}
where $N$ is the number of unlabeled source samples in a batch. $f\left(x_{i} ; \theta_t, \xi\right)$ and $f\left(x_{i} ; \theta_{t}, \xi^{\prime}\right)$ denote the outputs of the student model $G$ and the intra-domain teacher model $G_{intra}$, respectively.

\subsubsection{Inter-domain structural knowledge transfer}

Intuitively, the synthetic images contain the same underlying structures as the original ones in a different imaging modality, despite the fact that they may hold different visual appearances for the reason of the domain shift. In particular, the segmentation masks of the source image $x_{i}^s$ and its corresponding synthetic target-like image $x_{i}^{s\rightarrow t}$ should be identical,~\emph{i.e.}, $ y_{i}^s= y_{i}^{s\rightarrow t}$.
During the dual cycle alignment process, multiple paired images are generated,~\emph{i.e.}, $\{[x^s, x^{s\rightarrow t}]; [x^{s\rightarrow t \rightarrow s}, x^{s\rightarrow t}]; [x^{t \rightarrow s}, x^{t}]; [x^{t \rightarrow s}, x^{t \rightarrow s \rightarrow t}]\}$, these images exhibit rich cross-modality structural information to explore. Motivated by these findings, we propose an inter-domain teacher model $G_{inter}$ based on self-ensembling to maintain structural consistency between teacher and student model predictions by feeding the unlabeled pairs as inputs to the models in a parallel manner, allowing the student model to be better guided by the domain-shared structural knowledge. Following~\cite{vu2019advent}, we explore the structural knowledge based on weighted self-information and ensure that the entropy maps of the student model and the inter-domain teacher model are similar. Specifically, we apply Shannon entropy~\cite{shannon2001mathematical} to generate the entropy map, and the structural consistency loss between the student model and the inter-domain teacher model is defined as:
\begin{equation}
\label{eq:loss_con}
\mathcal{L}_{{con}}^{{inter}}=\frac{1}{N} \sum_{i=1}^{N} \frac{1}{H \times W} \sum_{v=1}^{\mathcal{V}}\left\|\mathbf{I}_{i, v}^{s}-\mathbf{I}_{i, v}^{t}\right\|^{2},
\end{equation}
where $N$ is the number of unlabeled input pairs in a batch,  $\mathcal{V}=\{1,2, \ldots, H \times W \}$ is the number of pixels in an image, and $\mathbf{I}_{i, v}^{s} = -\mathbf{p}_{i, v}^{s} \circ \log \mathbf{p}_{i, v}^{s}$ is the weighted self-information at the $v$-th pixel based on the $i$-th output from the student model. Similarly, $\mathbf{I}_{i, v}^{t}$ corresponds to that from the inter-domain teacher model.
The notation $\circ$ denotes the element-wise product,~\emph{i.e.}, Hadamard product, while $\log$ denotes the base-$2$ logarithmic equation.

\subsection{Self-ensembling adversarial learning}

Although we realize partial feature adaptation in dual teacher-student learning by explicitly enforcing both inter-domain and intra-domain consistency, more domain-invariant discriminative features should be aligned in the feature space.
Inspired by both adversarial learning~\cite{GAN} and self-ensembling learning~\cite{meanteacher}, we integrate adversarial learning into our self-ensembling teacher-student network in the semantic prediction space in a mutually beneficial manner. 
As depicted in the right-most block of Fig.~\ref{fig:archi}, we construct two discriminators,~\emph{i.e.}, an intra-domain discriminator $D_{intra}$ and an inter-domain discriminator $D_{inter}$ for self-ensembling adversarial learning. 
Intuitively, the intra-domain discriminator $D_{intra}$, the intra-domain teacher $G_{intra}$, and the student network $G$ naturally form the GANs, where $G_{intra}$ and $G$ play the role of generators while $D_{intra}$ aims to figure out whether the features are extracted from $G_{intra}$ or $G$. The discriminator $D_{intra}$ back-propagates the adversarial gradients with an adversarial loss to the student network $G$ in the feature spaces, which encourages $G$ to produce more similar feature maps as $G_{intra}$. The adversarial loss can be expressed as follows:
\begin{equation}
\begin{aligned}
\mathcal{L}_{adv}\left(G, D_{intra}\right)=& \mathbb{E}_{x \sim \mathcal{D}_{s}}\left[\log \left(1-D_{intra}\left(G\left(x\right)\right)\right)\right]+ \\
&\mathbb{E}_{x^{\prime} \sim \mathcal{D}_{s}^{s}}\left[\log D_{intra}\left(G_{intra}\left(x^{\prime}\right)\right)\right].
\end{aligned}
\end{equation}

Similarly, for the inter-domain discriminator $D_{inter}$, the adversarial loss can be formulated as:
\begin{equation}
\begin{aligned}
\mathcal{L}_{adv}\left(G, D_{inter}\right)=& \mathbb{E}_{x \sim \mathcal{D}_{s}}\left[\log \left(1-D_{inter}\left(G\left(x\right)\right)\right)\right]+ \\
&\mathbb{E}_{x^{\prime} \sim \mathcal{D}_{t}^{s}}\left[\log D_{inter}\left(G_{inter}\left(x^{\prime}\right)\right)\right].
\end{aligned}
\end{equation}

In the proposed self-ensembling adversarial learning scheme, the discriminator acts as a teaching assistant network to help the student learn domain-invariant features under self-ensembling consistency, improving the generalization capability. On the other hand, the consistency regularization behind self-ensemble learning prolongs the minimax competition between the discriminator and generators, improving the stability of the adversarial learning process. Similar to~\cite{tsai2018learning}, we introduce the multi-level deep supervision strategy~\cite{lee2015deeply} into lower layers for enhancing the alignment of low-level features. To elaborate, auxiliary classifiers are built with one 1 $\times$ 1 convolution followed by an upsampling layer to obtain multi-scale outputs from different layers. We connect auxiliary discriminators with the outputs of different layers to facilitate multi-level feature alignment. The multi-level adversarial losses $\mathcal{L}_{adv}^{intra}$ and $\mathcal{L}_{adv}^{inter}$ can be extended as:
\begin{equation}
\label{eq:adv}
\begin{aligned}
\mathcal{L}_{adv}^{intra} = \sum_{i} \lambda_{a d v}^{i} \mathcal{L}_{a d v}^{i}\left(G^i, D_{intra}^i\right),\\
\mathcal{L}_{adv}^{inter} = \sum_{i} \lambda_{a d v}^{i} \mathcal{L}_{a d v}^{i}\left(G^i, D_{inter}^i\right),
\end{aligned}
\end{equation}
where $i$ indicates the last $i$-th layer connected with the discriminator $D^i$, and $G^i$ generates the outputs of layer $i$.

\subsection{Student model and overall training strategies}
Under the multi-level supervision with the corresponding source labels $y^{s}$, the student model $G$ is trained on the labeled source domain $\mathcal{D}_{s}^l$ with the deeply supervised loss $\mathcal{L}_{{stu }}^{s e g}$, which is formulated as:
\begin{equation}
\label{eq:loss_stu}
\mathcal{L}_{{stu }}^{s e g}=\sum_{i} \lambda_{seg}^{i} \left[\mathcal{L}_{ {ce }}\left(y^{s}, G\left(x^{s}\right)\right)+\mathcal{L}_{ {dice}}\left(y^{s}, G\left(x^{s}\right)\right)\right],
\end{equation}
where $\lambda_{seg}$ controls the strengths of segmentation losses from different layers, and $\mathcal{L}_{ {ce }}$ and $\mathcal{L}_{ {dice }}$ are the cross-entropy loss and Dice loss, respectively. Ultimately, we integrate losses from Eqs.~(\ref{eq:loss_kd}), (\ref{eq:loss_con}), (\ref{eq:adv}) and (\ref{eq:loss_stu}) together to obtain the final training objective of the student network, which is written as:
\begin{equation}
\label{eq:total}
\begin{array}{r}
\mathcal{L}_{stu}=\underbrace{\mathcal{L}_{s t u}^{seg}}_{\text {supervised}} + \underbrace{\lambda_{con}^{intra}\mathcal{L}_{con}^{intra}+\lambda_{adv}^{intra}\mathcal{L}_{adv}^{intra}}_{\text {intra-domain}} +  \\
\underbrace{\lambda_{con}^{inter} \mathcal{L}_{con}^{inter}+ \lambda_{adv}^{inter}\mathcal{L}_{adv}^{inter}}_{\text {inter-domain}},
\end{array}
\end{equation}
where $\lambda_{con}^{intra}$, $\lambda_{adv}^{intra}$, $\lambda_{con}^{inter}$ and $\lambda_{adv}^{inter}$ are the weights of the associated losses. With the proposed method, the student model can effectively distill multi-domain knowledge from different domains for efficient cross-domain image segmentation.

\section{Experiments}
\label{sec:experiments}

\subsection{Dataset}
To validate the effectiveness of the proposed~\ourmethod~framework, we employ two popular medical image segmentation tasks, namely cardiac substructure segmentation and abdominal multi-organ segmentation. Unpaired MRI and CT images from different subjects are collected for both tasks. We conduct extensive experiments to validate the proposed~\ourmethod~for UDA with different directions,~\emph{e.g.}, $\text{MRI} \rightarrow \text{CT}$ and different source label ratios,~\emph{e.g.}, $25\%$.

\subsubsection{Cardiac substructure segmentation} 
For cross-modality cardiac segmentation, we used the Multi-Modality Whole Heart Segmentation (MM-WHS) Challenge 2017 dataset~\cite{Zhuang2016Multi}, consisting of unpaired $20$ CT and $20$ MRI scans with ground-truth pixel-level annotations acquired in the real clinical environment. Following~\cite{chen2020unsupervised}, the network was developed to segment four individual cardiac substructures,~\emph{i.e.}, myocardium of the left ventricle (MYO), left atrium blood cavity (LAC), left ventricle blood cavity (LVC), and ascending aorta (AA).

\subsubsection{Abdominal multi-organ segmentation}
To further assess the generalizability of the proposed~\ourmethod, we employed our framework on another challenging task, abdominal multi-organ segmentation for UDA. We used the training dataset comprised of $30$ abdominal CT images from the MICCAI $2015$ Multi-Atlas Abdomen Labeling Challenge and $20$ T2-SPIR MRI images from the ISBI $2019$ CHAOS Challenge~\cite{kavur2021chaos} for UDA. Four abdominal organs were included for segmentation,~\emph{i.e.}, liver, right kidney~(R. kidney), left kidney~(L. kidney), and spleen.

Following the universal UDA setting as defined in~\cite{chen2020unsupervised}, for both tasks, we randomly split each modality with $80\%$ subjects for training and $20\%$ subjects for testing. The source label ratio was set $25\%$ to simulate the source label scarcity scenario, where $25 \%$ labeled training data were randomly selected for training. Since target labels are inaccessible in the UDA scenario, we utilized the testing set on the source domain for hyperparameter optimization.
For preprocessing, the original MRI and CT scans possess varying fields of view in both datasets. To map the view of multi-modal images to the same scene, similar to~\cite{chen2020unsupervised}, for the cardiac dataset, we cropped the slices centered on the heart area in the coronal plane into the size of $256 \times 256$. While for the abdominal images, we generated a three-dimensional bounding box for each volume to filter out slices without the four abdominal organs on the axial plane. With the $z$-score transformation, all volumes were normalized to a mean of zero and a standard deviation of one.
To avoid overfitting, we used online data augmentations randomly, including rotation, scaling, and shear transformation during the training phase.

\subsubsection{Evaluation metrics}
To compare the segmentation performances of different approaches, two widely-used metrics in segmentation tasks,~namely the Dice coefficient (Dice) and the average symmetric surface distance (ASD) were used in our experiments. 
The Dice quantifies the accuracy of voxel-wise segmentation between the predictions and the ground truth masks, with a higher Dice value indicating better segmentation results. 
While ASD estimates the average distances between the surface of the ground truth masks and the predictions in three dimensions, with a lower ASD value indicating better performance at boundaries. Following the settings in previous work~\cite{chen2019synergistic,chen2020unsupervised}, we evaluated different methods based on the voxel-wise prediction masks at the subject level, and both metrics are presented with the across-subject means and variances of the segmentation results, formatted as \textit{mean} (\textit{std}).

\subsubsection{Implementation details}
In general, the whole model is end-to-end trainable, but this would require a large GPU memory. For computational efficiency, we first closely followed the original CycleGAN settings~\cite{zhu2017unpaired} to optimize the dual cycle alignment module for image alignment and complementary data generation. Then, we jointly trained the dual-teacher network for dual-domain knowledge transfer and performed self-ensembling adversarial learning with the objective in Eq. (\ref{eq:total}). 
Random perturbations (\emph{i.e.,} Gaussian noise) were applied in the network inputs.  It is noted that the student model in our method was trained on the source domain under source label scarcity. Therefore, following~\cite{zhang2018task}, we evaluated our model on the transformed source-like images $x^{t \rightarrow s}$ instead of target domain images $x$.
In our framework, U-Net~\cite{ronneberger2015u} was implemented as the network backbone for both the student model $G$ and the teacher models $G_{inter}$, $G_{intra}$. For discriminators $D_{inter}$ and $D_{intra}$, similar to~\cite{tsai2018learning,chen2019synergistic},  we implemented a fully convolutional network with five convolutional~(Conv) layers having a kernel size of $4 \times 4$ and a stride of $2$. The number of feature channels for each of the $5$ layers is $\{64, 128, 256, 512, 1\}$, respectively. Each Conv layer is followed by a leaky ReLU parameterized by $0.2$, with the exception of the final layer.
We trained the framework using the Adam optimizer with a momentum of $0.9$ for $150$ epochs on one NVIDIA Tesla V100 GPU with 32GB memory. The learning rate was linearly warmed up during the first $30$ epochs from $0$ to $0.005$ to reduce volatility in the early training stages. The total batch size was $16$, containing $8$ labeled samples for supervised learning and $8$ unlabeled data pairs for teacher-student training in each batch. The weights $\{\lambda_{con}^{intra}, \lambda_{adv}^{intra}, \lambda_{con}^{inter}, \lambda_{adv}^{inter}\}$ ramped up individually from $0$ to their maximum values $ L_{max} = \{1, 0.01, 0.1, 0.01\}$ with a sigmoid function $\lambda(t)= L_{max} * e^{\left(-5\left(1-t / t_{\max }\right)^{2}\right)}$ to maintain the balance between different losses, where $t$ and $t_{max}$ represent the current and maximum epochs. For multi-level self-ensembling adversarial learning, we set layers to be connected $i=\{1,3\}$, and then connected the last block and the third last block of the decoding stage of U-Net to classifiers and discriminators. Accordingly, we empirically set $\lambda_{adv}^i = \{1,0.1\}$ and $\lambda_{seg}^i = \{1,0.1\}$.

\subsection{Performance comparison with other methods}
To access the effectiveness of the~\ourmethod~framework, we extensively compare our method with representative UDA methods,~\emph{i.e.}, ADDA~\cite{tzeng2017adversarial}, CycleGAN~\cite{zhu2017unpaired}, and SIFA~\cite{chen2020unsupervised}. ADDA and CycleGAN are well-established methods for feature adaptation and image adaptation, respectively. While the SIFA framework conducts both image and feature adaptation. For a fair comparison, we used the same U-Net segmentation backbone as MT-UDA and LE-UDA for the implementation of ADDA and CycleGAN. For ADDA, we first adopted the hyperparameter settings in the official implementation\footnote{https://github.com/jhoffman/cycada\_release} and fine-tuned hyperparameters by observing losses curves of generators and discriminators in the feature adaptation process for stable adversarial learning. For CycleGAN, we followed the official implementation\footnote{https://github.com/junyanz/pytorch-CycleGAN-and-pix2pix} for the configurations and hyperparameter settings, and the segmentation model was optimized on the transformed target-like images.
For SIFA, we followed the official implementation for the configurations of network architectures\footnote{https://github.com/cchen-cc/SIFA}. For baselines, we trained a U-Net model with only labeled images from the source domain without any adaptation techniques and applied the model directly to the target domain, referred to as ``W/o adaptation” lower bound, and trained another U-Net model with only labeled target domain data, named ``Supervised-only” upper bound. For the ``W/o adaptation” lower bound, the model was optimized on the source domain, while the model was optimized on the target domain for the ``Supervised-only” upper bound.

\subsubsection{Cardiac substructure segmentation}
Table~\ref{tab:mmwhs_CT} demonstrates the performance comparison of various approaches for cross-modality segmentation~(MRI → CT) on the cardiac dataset under different source label ratios. 
On the full-labeled source dataset, the ``W/o adaptation” lower bound obtained the average Dice of $17.6\%$ on CT images, while the ``Supervised-only” upper bound obtained the average Dice of $86.6\%$. 
The large performance difference between the lower and upper bounds demonstrates the severe domain gap across modalities. 
With the help of unlabeled target data, the UDA methods improved the performance at different levels by using either image adaptation or feature adaptation.
Although SIFA achieved promising adaptation performance under domain shift, demonstrating the effectiveness of concurrent adaptation from different aspects, these UDA methods in general offer a severely degraded performance when subjected to a smaller source label ratio. The reason is that annotation scarcity limits the segmentation performance within the source domain and then restricts the cross-domain adaptation performance. 
In comparison to other existing methods, MT-UDA has a stable segmentation performance with a smaller performance drop as the source label ratio decreases, achieving higher mean Dice and lower ASD with limited source labels as compared to other existing UDA methods. \ourmethod~further increased the adaptation performance to $70.8\%$ in mean Dice, which is the closest to the ``Supervised-only” upper bound with $25\%$ annotations.
We hypothesize that there could be two reasons for this. First, we generate multiple different intermediate domains, with rich inter-domain and intra-domain knowledge to exploit for addressing both the domain shift and label scarcity problems.  Second, our method employs self-ensembling models to fully utilize all available data sources from different domains, thereby leveraging multi-domain knowledge effectively for better UDA performance under source label scarcity. Moreover, the proposed dual self-ensembling adversarial learning strategy can provide additional assistance on the feature level to facilitate knowledge transfer and domain adaptation, yielding an improvement of 3\% in mean Dice as compared to our prior work, MT-UDA.

\begin{figure*}[tbp]
    \centering
    \includegraphics[width =0.9\linewidth]{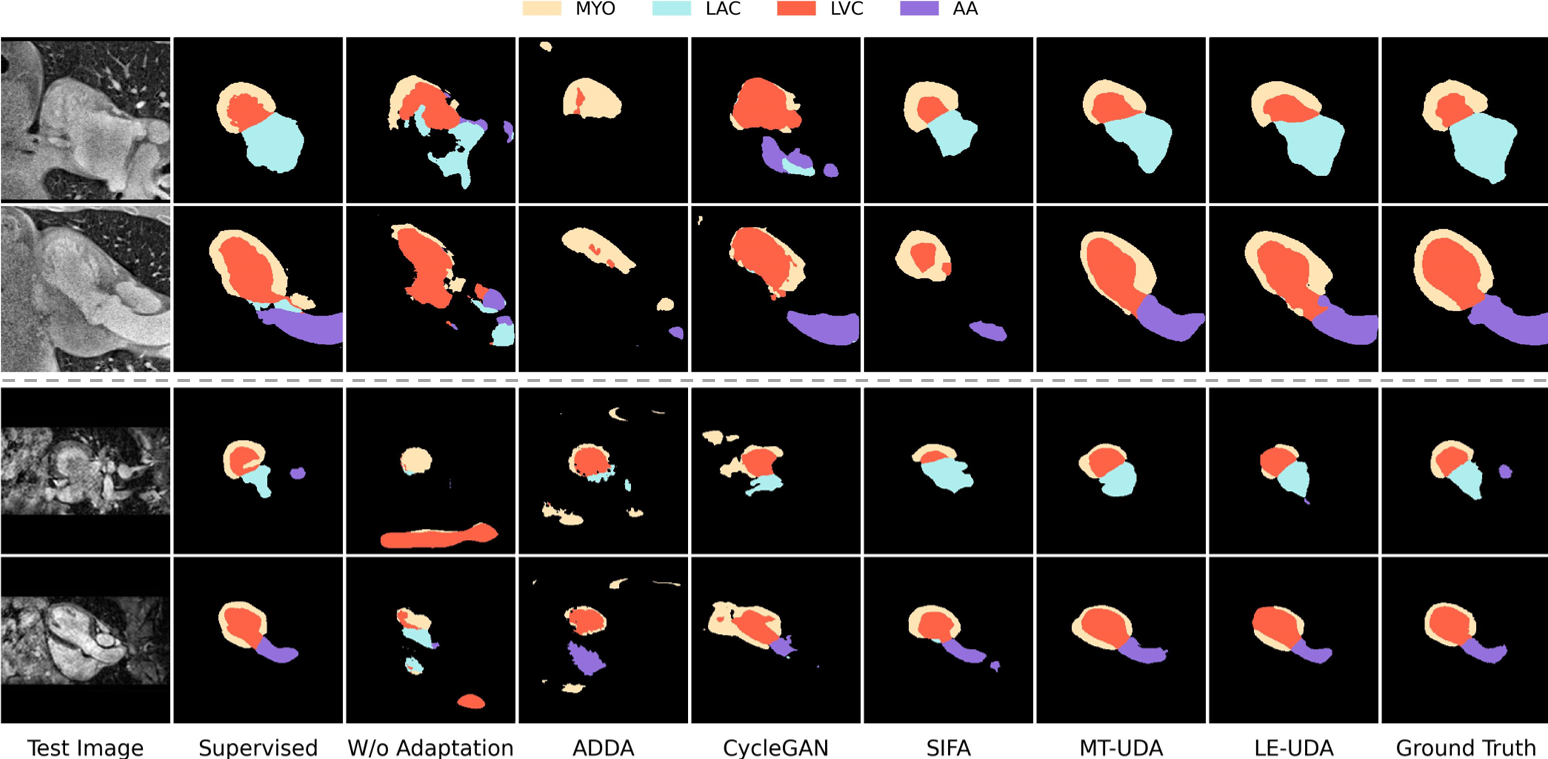}
    \caption{Visual comparison of segmentation results generated by various approaches using 25\% source labels for cardiac CT images (first two rows) and cardiac MRI images (last two rows).}
    \label{fig:mmwhs}
\end{figure*}

To further demonstrate the feasibility of our method on bidirectional cross-modality adaptation, we reversed the adaptation direction with CT serving as the source domain and MR serving as the target domain,~\emph{i.e.,} CT → MRI. Table~\ref{tab:mmwhs_MR} presents the segmentation performance on cardiac MR images. Trained with limited labeled CT data,  the ``W/o adaptation” lower bound achieved only $9.2\%$ in mean Dice and $24.8$ in mean ASD. By leveraging unlabeled MRI data, the best UDA method,~\emph{i.e.,} SIFA would greatly improve the performance with $49.8\%$ in mean Dice and $8.2$ in mean ASD, however, the performance is still heavily limited by the sparse source annotations, which is far below SIFA on full source annotation. By exploiting both unlabeled MR and CT data, both MT-UDA and LE-UDA yielded higher mean Dice and lower mean ASD over SIFA under different source label ratios, and LE-UDA achieved better performance than our prior work MT-UDA, indicating the effectiveness of LE-UDA on bidirectional unsupervised domain adaptation. The qualitative segmentation results of various approaches using $25\%$ annotations are displayed in Fig.~\ref{fig:mmwhs}. We observe that the ``W/o adaptation” lower bound cannot obtain correct and meaningful predictions for any cardiac substructures due to the domain shift. Compared to other UDA methods, our model is capable of identifying different substructures with clean and smooth boundaries, producing more reliable masks with fewer false positives in either the adaptation from MRI to CT or CT to MRI.

\begin{table*}[t]
\centering
\setlength\tabcolsep{5pt}
\caption{Performance comparison between different approaches for CT cardiac substructure segmentation under different source label ratios. The metrics are presented in the format of mean (std).}
\scalebox{0.9}{
\begin{tabular}{c|c|ccccc|ccccc} 
\toprule
\multicolumn{12}{c}{Cardiac MRI → Cardiac CT}                                                                                                                                            \\ 
\hline
\multirow{2}{*}{Method} & \multirow{2}{*}{Label Ratio (\%)} & \multicolumn{5}{c|}{Dice}                                    & \multicolumn{5}{c}{ASD}                                     \\ 
\cline{3-12}
                        &                                   & AA         & LAC        & LVC        & MYO         & Average & AA         & LAC        & LVC       & MYO        & Average  \\ 
\hline
Supervised-only     & 100                                & 85.5(13.2) & 88.6(3.4)  & 88.1(4.9)  & 84.2(5.6)   & 86.6    & 2.1(1.2)   & 8.9(3.8)   & 4.1(2.7)  & 2.7(1.5)   & 4.4      \\
W/o adaptation          & 100                                & 37.4(28.0) & 26.8(18.9) & 2.5(3.1)   & 3.6(3.9)    & 17.6    & 42.6(23.3) & 34.6(3.3)  & N/A       & 29.2(15.2) & N/A      \\
\hline
ADDA~\cite{tzeng2017adversarial}               & 100          & 44.4(13.3) & 28.5(15.5) & 39.7(13.2) & 26.7(12.3)  & 34.8    & 36.5(5.9)  & 26.6(9.8)) & 23.2(7.7) & 35.4(6.6)  & 30.4     \\
CycleGAN~\cite{zhu2017unpaired}          & 100          & 55.5(7.7)  & 57.9(7.6)  & 53.1(9.3)  & 30.2(22.3)  & 49.2    & 21.1(2.0)  & 20.0(7.1)  & 18.2(7.1) & 12.7(3.0)  & 18.0       \\
SIFA~\cite{chen2020unsupervised}              & 100          & 78.3(3.0)  & 77.5(5.3)  & 73.1(8.6)  & 61.2(7.9)   & 72.5    & 9.3(1.6)   & 8.7(3.7)   & 7.0(2.4)  & 6.9(2.1)   & 8.0        \\
MT-UDA~\cite{zhao2021mt}             & 100          & 73.1(5.8)  & 82.1(4.1)  & 72.8(13.9) & 61.9(11.4)  & 72.5    & 22.7(4.9)  & 12.2(3.7)  & 11.2(4.7) & 8.2(2.8)   & 13.6     \\
LE-UDA (Ours)       & 100          & 72.9(4.7)  & 83.7(3.0)  & 74.6(13.5) & 62.1(12.4)  & 73.3    & 23.0(3.9)  & 5.8(1.6)   & 6.6(3.1)  & 6.6(3.1)   & 10.4     \\ 
\hline\hline
Supervised-only &   25      & 82.8(11.3)  &83.8(3.4)  &74.8(10.1) &61.0(17.7)  & 75.6  & 6.8(4.1)   & 11.3(1.4) & 7.6(2.0)  & 8.9(2.7)  &  8.6   \\
W/o adaptation      & 25           & 14.9(12.1) & 4.1(4.1)   & 18.5(14.4) & 10.7(11.2) & 12.0    & 52.0(8.8)  & 36.8(9.0)  & 38.9(8.7) & 38.1(14.7) & 41.5     \\
\hline
ADDA~\cite{tzeng2017adversarial}               & 25           & 31.7(5.7)  & 15.4(7.1)  & 9.0(6.3)   & 31.2(6.1)   & 21.8    & 39.6(6.3)  & 27.4(14.0) & 28.0(8.9) & 22.0(4.2)  & 29.2     \\
CycleGAN~\cite{zhu2017unpaired}           & 25           & 31.7(16.6) & 22.7(16.4) & 58.1(25.4) & 31.9(23.3)  & 36.1    & 19.2(9.8)  & 24.5(9.7)  & 24.5(9.7) & 20.4(27.5) & 19.1     \\
SIFA~\cite{chen2020unsupervised}              & 25           & 42.3(17.4) & 58.9(7.0)  & 47.4(20.4) & 44.7(18.1)  & 48.3    & 10.3(3.4)  & 14.4(8.4)  & 9.3(4.6)  & 6.9(2.7)   & 10.2     \\
MT-UDA~\cite{zhao2021mt}              & 25           & 67.1(6.5)  & 78.0(5.4)  & \textbf{71.1(8.1)}  & 55.0(11.4)  & 67.8    & \textbf{6.6(2.2)}   & 11.6(6.9)  & 9.2(3.5)  & \textbf{10.4(3.3)}  & \textbf{9.5}      \\
LE-UDA (Ours)       & 25           & \textbf{76.7(6.3)}  & \textbf{80.9(4.6)}  & 67.3(12.1) & \textbf{58.4(12.3)}  & \textbf{70.8}    & 7.3(2.3)   & \textbf{10.8(5.9)}  & \textbf{9.2(3.2)}  & 11.0(5.4)  & 9.6      \\
\bottomrule
\end{tabular}
}
\label{tab:mmwhs_CT}
\end{table*}

\begin{table*}[t]
\centering
\setlength\tabcolsep{5pt}
\caption{Performance comparison between different approaches for MRI cardiac substructure segmentation under different source label ratios. The metrics are presented in the format of mean (std).}
\scalebox{0.9}{
\begin{tabular}{c|c|ccccc|ccccc} 
\toprule
\multicolumn{12}{c}{Cardiac CT → Cardiac MRI}                                                                                                                                            \\ 
\hline
\multirow{2}{*}{Method} & \multirow{2}{*}{Label Ratio (\%)} & \multicolumn{5}{c|}{Dice}                                    & \multicolumn{5}{c}{ASD}                                     \\ 
\cline{3-12}
                        &                                   & AA         & LAC        & LVC        & MYO        & Average & AA        & LAC        & LVC        & MYO        & Average  \\ 
\hline
Supervised-only         & 100                               & 84.2(3.4)  & 82.8(8.6)  & 92.3(1.7)  & 80.4(3.8)  & 84.9    & 2.9(1.7)  & 3.4(1.8)   & 1.6(0.6)   & 2.0(0.8)   & 2.5      \\
W/o adaptation          & 100                               & 24.1(6.5)  & 16.5(5.5)  & 28.6(21.3) & 7.8(7.8)   & 19.3    & 16.0(1.5) & 19.8(2.9)  & 14.7(7.6)  & 12.6(6.1)  & 15.8     \\ 
\hline
ADDA~                   & 100                               & 12.1(8.8)  & 15.8(4.8)  & 56.1(18.3) & 25.2(9.4)  & 27.3    & 25.9(7.1) & 24.1(2.0)  & 12.6(6.1)  & 14.4(5.3)  & 19.3     \\
CycleGAN~               & 100                               & 60.4(6.5)  & 55.8(10.7) & 57.7(8.8)  & 28.4(8.6)  & 50.6    & 7.1(2.8)  & 7.0(4.0)   & 7.0(4.0)   & 13.4(6.8)  & 9.3      \\
SIFA~                   & 100                               & 65.3(4.9)  & 62.3(12.8) & 78.9(5.1)  & 47.3(7.2)  & 63.4    & 7.3(3.5)  & 7.4(1.8)   & 3.8(0.8)   & 4.4(0.8)   & 5.7      \\
MT-UDA~                 & 100                               & 70.3(7.8)  & 73.4(7.1)  & 76.4(25.9) & 42.4(1.4)  & 65.6    & 4.7(2.7)  & 2.9(0.6)   & 2.9(2.3)   & 4.5(3.7)   & 3.7      \\
LE-UDA (Ours)           & 100                               & 71.3(5.3)  & 69.6(10.2) & 79.1(22.0) & 59.4(15.7) & 69.8    & 3.9(1.7)  & 3.8(1.3)   & 2.5(1.7)   & 3.7(1.7)   & 3.5      \\ 
\hline\hline
Supervised-only         & 25                                & 71.3(5.4)  & 64.1(15.3) & 84.7(7.7)  & 67.8(8.8)  & 72.0    & 7.8(3.1)  & 10.0(8.2)  & 3.4(2.0)   & 6.2(5.5)   & 6.8      \\
W/o adaptation          & 25                                & 1.7(1.8)   & 12.9(9.5)  & 20.7(10.8) & 1.3(0.6)   & 9.2     & 28.9(8.3) & 22.7(2.4)  & 29.7(13.4) & 18.0(3.2)  & 24.8     \\ 
\hline
ADDA~                   & 25                                & 14.1(14.6) & 26.3(12.2) & 48.2(17.2) & 11.3(3.3)  & 25.0    & 22.2(2.9) & 24.2(3.9)  & 14.0(2.1)  & 23.4(5.7)  & 20.9     \\
CycleGAN~               & 25                                & 45.3(9.8)  & 12.5(10.2) & 55.0(30.9) & 25.6(15.3) & 34.6    & 9.4(4.2)  & 24.7(11.0) & 12.7(9.9)  & 11.8(6.7)  & 14.6     \\
SIFA~                   & 25                                & 56.9(2.9)  & 53.2(15.9) & 54.5(25.2) & 34.7(14.5) & 49.8    & 9.7(3.5)  & 6.1(2.7)   & 7.9(5.8)   & 9.0(6.7)   & 8.2      \\
MT-UDA~                 & 25                                & 70.1(4.4)  & 66.7(12.1) & 74.2(27.6) & \textbf{39.8(25.2)} & 62.7    & 5.0(2.9)  & 3.9(1.9)   & 3.4(2.8)   & 10.2(12.3) & 5.6      \\
LE-UDA (Ours)           & 25                                & \textbf{74.1(6.6)}  & \textbf{72.0(8.5)}  & \textbf{84.1(8.9)}  & 35.9(13.6) & \textbf{66.5}    & \textbf{4.7(3.5) } & \textbf{2.7(0.5)}   & \textbf{2.4(0.8)}   & \textbf{6.1(2.3)}   & \textbf{4.0}     \\
\bottomrule
\end{tabular}
}
\label{tab:mmwhs_MR}
\end{table*}

\begin{table*}[tbp]
\centering
\setlength\tabcolsep{5pt}
\caption{Performance comparison between different approaches for CT abdominal multi-organ segmentation under different source label ratios. The metrics are presented in the format of mean (std).}
\scalebox{0.9}{
\begin{tabular}{c|c|ccccc|ccccc} 
\toprule
\multicolumn{12}{c}{Abdominal MRI → Abdominal CT}                                                                                                                                            \\ 
\hline
\multirow{2}{*}{Method} & \multirow{2}{*}{Label Ratio (\%)} & \multicolumn{5}{c|}{Dice}                                    & \multicolumn{5}{c}{ASD}                                     \\ 
\cline{3-12}
                        &                                   & Liver      & R. kidney  & L. kidney  & Spleen     & Average & Liver     & L. kidney  & R. kidney & Spleen   & Average  \\ 
\hline
Supervised-only         & 100                               & 92.1(5.2)  & 82.8(7.2)  & 85.8(4.3)  & 93.7(3.0)  & 88.6    & 2.2(2.4)  & 1.8(1.3)   & 1.1(0.6)  & 1.0(0.6) & 1.5      \\
W/o adaptation          & 100                               & 57.7(21.3) & 47.1(13.4) & 5.6(16.9)  & 61.9(9.1)  & 53.1    & 5.8(2.4)  & 10.5(4.7)  & 13.0(3.3) & 5.8(2.7) & 8.8      \\ 
\hline
ADDA~                   & 100                               & 67.0(18.6) & 73.0(6.5)  & 67.0(17.3) & 81.0(7.1)  & 72.0    & 7.8(4.6)  & 8.9(5.1)   & 5.2(2.0)  & 5.5(2.4) & 6.9      \\
CycleGAN~               & 100                               & 73.4(12.7) & 82.7(2.6)  & 79.9(3.8)  & 79.3(6.6)  & 78.8    & 2.6(0.9)  & 1.5(0.4)   & 2.0(0.3)  & 3.5(1.8) & 2.4      \\
SIFA~                   & 100                               & 79.4(12.7) & 83.1(8.7)  & 85.7(3.7)  & 84.7(2.8)  & 83.2    & 1.7(0.8)  & 1.4(0.7)   & 1.7(0.5)  & 2.6(1.0) & 1.9      \\
MT-UDA~                 & 100                               & 78.5(12.5) & 85.1(4.1)  & 85.3(7.0)  & 87.3(2.9)  & 84.0    & 2.3(1.6)  & 1.3(0.7)   & 2.1(1.7)  & 1.9(1.0) & 1.9      \\
LE-UDA (Ours)           & 100                               & 80.7(9.6)  & 84.5(5.9)  & 84.5(10.0) & 87.8(3.3)  & 84.4    & 1.8(1.2)  & 1.0(0.4)   & 2.0(1.9)  & 1.8(1.2) & 1.6      \\ 
\hline\hline
Supervised-only         & 25                                & 88.1(10.0) & 76.1(17.5) & 84.5(7.2)  & 94.4(3.0)  & 85.8    & 1.0(0.5)  & 1.6(1.5)   & 1.6(0.6)  & 1.5(1.1) & 1.4      \\
W/o adaptation          & 25                                & 46.0(26.8) & 16.5(9.0)  & 25.2(5.7)  & 64.2(8.0)  & 38.0    & 23.3(7.7) & 20.8(10.4) & 26.1(4.8) & 8.1(3.3) & 19.6     \\ 
\hline
ADDA~                   & 25                                & 52.1(25.9) & 39.8(11.5) & 34.7(21.1) & 67.9(14.9) & 48.6    & 24.3(9.0) & 14.4(6.1)  & 11.4(5.9) & 5.8(3.2) & 14.0       \\
CycleGAN~               & 25                                & 51.7(24.5) & 67.6(5.8)  & 64.4(14.2) & 77.8(9.0)  & 65.4    & 10.6(6.0) & 2.2(0.9)   & 3.7(1.6)  & 4.1(2.0) & 5.2      \\
SIFA~                   & 25                                & 61.4(21.9) & 77.9(8.7)  & 77.6(6.2)  & 78.1(6.2)  & 73.8    & 7.5(8.4)  & 6.7(5.6)   & 1.9(0.4)  & 4.1(2.9) & 5.0      \\
MT-UDA~                 & 25                                & 78.3(11.1) & 79.7(5.3)  & \textbf{79.4(8.8)}  & 83.3(2.9)  & 80.2    & 1.8(1.4)  & \textbf{2.5(1.7)}   & 4.0(1.8)  & 3.2(1.6) & 2.9      \\
LE-UDA (Ours)           & 25                                & \textbf{80.5(9.8)}  & \textbf{82.4(7.1)}  & 78.9(12.3) & \textbf{89.3(2.6)} & \textbf{82.8}   & \textbf{1.6(1.1)}  & 2.9(2.1)   &\textbf{ 2.5(1.9)}  & \textbf{2.0(1.4)} & \textbf{2.3}     \\
\bottomrule
\end{tabular}
}
\label{tab:abdominal_CT}
\end{table*}

\begin{table*}[tbp]
\centering
\setlength\tabcolsep{5.4pt}
\caption{Performance comparison between different approaches for MRI abdominal multi-organ segmentation under different source label ratios. The metrics are presented in the format of mean (std).}
\scalebox{0.9}{
\begin{tabular}{c|c|ccccc|ccccc} 
\toprule
\multicolumn{12}{c}{Abdominal CT → Abdominal MRI}                                                                                                                                   \\ 
\hline
\multirow{2}{*}{Method} & \multirow{2}{*}{ Label Ratio (\%)} & \multicolumn{5}{c|}{Dice}                                  & \multicolumn{5}{c}{ASD}                                 \\ 
\cline{3-12}
                        &                                   & Liver              & R. kidney          & L. kidney          & Spleen             & Average       & Liver             & L. kidney         & R. kidney         & Spleen            & Average       \\ 
\hline
Supervised-only         & 100                               & 90.9(7.3)          & 89.7(5.9)          & 91.6(2.4)          & 93.8(6)            & 91.5          & 1.0(1.3)          & 1.1(1.3)          & 1.1(0.7)          & 2.9(5.1)          & 1.5           \\
W/o adaptation          & 100                               & 78.0(3.8)          & 48.4(21)           & 51.6(15.4)         & 46.5(21.3)         & 56.1          & 1.5(0.5)          & 3.0(1.2)          & 2.6(0.8)          & 4(1.4)            & 2.8           \\ 
\hline
ADDA~                   & 100                               & 81.2(5.1)          & 80.7(6.3)          & 68.2(13.3)         & 58.4(16.4)         & 72.1          & 3.3(1)            & 2.9(1.7)          & 2(0.7)            & 3.8(1.7)          & 3.0           \\
CycleGAN~               & 100                               & 83.7(3.8)          & 80.0(2.8)          & 80.7(4.2)          & 86.6(3.7)          & 82.7          & 3.9(4.6)          & 1.2(0.4)          & 2.3(2.6)          & 1.3(0.3)          & 2.2           \\
SIFA~                   & 100                               & 89.8(4.4)          & 87.4(2.9)          & 79.3(7.0)          & 86.3(3.9)          & 85.7          & 1.5(1.7)          & 1.5(1.7)          & 1.0(0.3)          & 1.6(0.6)          & 1.2           \\
MT-UDA~                 & 100                               & 91.1(2.9)          & 89.2(3.0)          & 87.3(2.9)          & 85.2(4.5)          & 88.2          & 0.5(0.2)          & 1.2(0.8)          & 0.8(0.3)          & 2.1(0.6)          & 1.1           \\
LE-UDA (Ours)           & 100                               & 90.8(3.7)          & 89.5(2.0)          & 89.3(2.7)          & 84.9(4.6)          & 88.6          & 0.6(0.3)          & 0.9(0.6)          & 0.5(0.1)          & 1.9(0.5)          & 1.0           \\ 
\hline\hline
Supervised-only         & 25                                & 81.8(6.9)          & 87.9(2.4)          & 87.9(2.4)          & 76.8(19.5)         & 82.7          & 2.5(2.4)          & 0.9(0.6)          & 1.0(0.3)          & 2.5(1.3)          & 1.7           \\
W/o adaptation          & 25                                & 78.2(7.5)          & 50.3(14.8)         & 23.1(8.0)          & 4.6(5.5)           & 39.1          & 2.0(0.7)          & 5.4(3.2)          & 6.5(3)            & 18.3(8.2)         & 8.1           \\ 
\hline
ADDA~                   & 25                                & 76.4(6.8)          & 68.9(10)           & 43.8(19.3)         & 17.4(11)           & 51.6          & 3.3(1.6)          & 5.6(2.5)          & 5.0(3)            & 11.7(5.2)         & 6.4           \\
CycleGAN~               & 25                                & 67.7(9.1)          & 74.4(2.6)          & 73.6(3.0)          & 85.5(0.7)          & 75.3          & 2.4(0.7)          & 3.1(2.4)          & 1.2(0.2)          & 1.6(0.3)          & 2.1           \\
SIFA~                   & 25                                & 84.4(7.6)          & 84.6(2.6)          & 56.9(18.7)         & 75.7(4.1)          & 75.4          & 1.2(0.9)          & 1.1(0.7)          & 1.5(0.8)          & 1.8(0.2)          & 1.4           \\
MT-UDA~                 & 25                                & 86.3(2.7)          & 89.1(2.1)          & 82.8(1.9)          & \textbf{83.9(5.1)} & 85.5          & 4.3(2.1)          & 0.9(0.5)          & 1.7(0.6)          & \textbf{2.0(0.6)} & 2.2           \\
LE-UDA (Ours)           & 25                                & \textbf{90.3(4.4)} & \textbf{90.3(4.4)} & \textbf{86.4(2.8)} & 82.3(5.6)          & \textbf{87.7} & \textbf{0.6(0.3)} & \textbf{0.4(0.1)} & \textbf{0.8(0.3)} & 2.1(0.7)          & \textbf{1.0}  \\
\bottomrule
\end{tabular}
}
\label{tab:abdominal_MR}
\end{table*}

\begin{figure*}[tbp]
    \centering
    \setlength\tabcolsep{5pt}
    \includegraphics[width =0.9\linewidth]{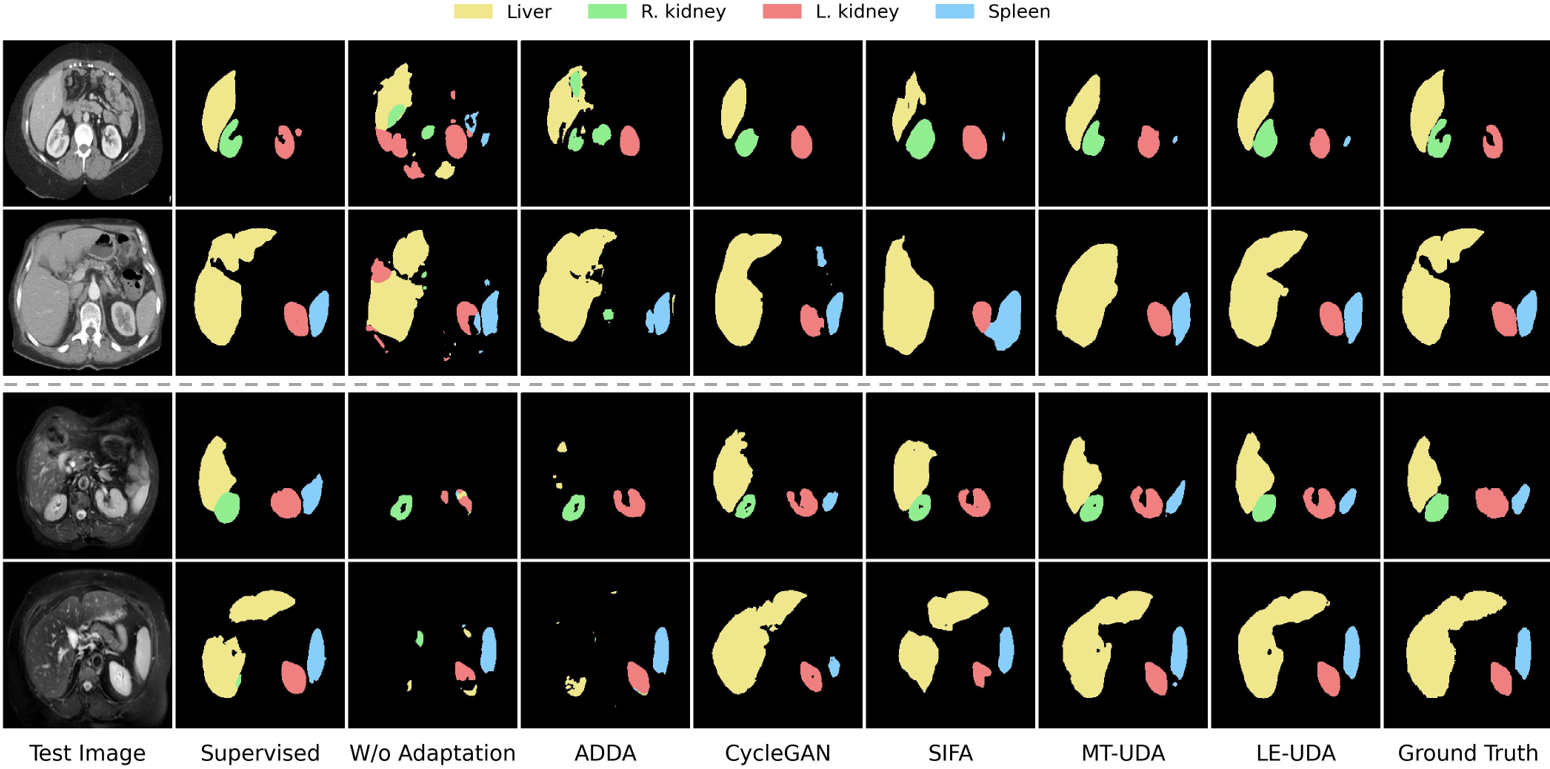}
    \caption{Visual comparison of segmentation results generated by various approaches using 25\% source labels for abdominal CT images (first two rows) and abdominal MRI images (last two rows).}
    \label{fig:abdom}
\end{figure*}
\subsubsection{Abdominal multi-organ segmentation} 
To further validate the proposed~\ourmethod~framework, we present the cross-modality segmentation results in both adaptation directions on abdominal images, as shown in Table~\ref{tab:abdominal_CT} and Table~\ref{tab:abdominal_MR}. Without domain adaptation, the model trained on $25\%$ labeled MRI images only achieved $38\%$ in mean Dice for segmenting CT images, and $39.1\%$ in mean Dice for the opposite direction. It is important to note that there exists a more severe performance gap between the ``W/o adaptation” lower bound and the ``Supervised-only” upper bound under source label scarcity than full annotation, which highlights the difficulty in unsupervised domain adaptation when subjected to source label scarcity. Like the previous cardiac dataset, we can see that the adaptation performances of different UDA strategies are heavily limited by source label scarcity. It is noted that feature adaptation (\emph{e.g.,} ADDA) has a more severe reduction in adaptation performance compared to image adaptation (\emph{e.g.,} CycleGAN). Image augmentation helps to mitigate the issue of label scarcity to some extent. Conversely, when performing feature adaptation, features in the sparse latent space are difficult to align. In this regard, although SIFA can achieve promising performance against domain shift, the adaptation performance is heavily limited when using $25\%$ annotations.
In contrast, our method can effectively leverage different types of data to obtain robust adaptation performance with limited source labels. Remarkably, for CT to MRI adaptation, our method even breaks the limitations of label scarcity and achieves superior performance compared to the ``Supervised-only” upper bound, which was trained with $25\%$ target labels. Due to inconsistency in the number of training data from different modalities,~\emph{i.e.,} CT~(30) and MRI~(20), we further trained the ``Supervised-only” upper bound with $6$ labeled MRI data for a more fair comparison, and the mean Dice is $88.5\%$, while our method achieves $87.7\%$, which is comparable. The visual comparisons on abdominal MRI and CT segmentation results in Fig.~\ref{fig:abdom} show the effectiveness of~\ourmethod~as well. For both MRI and CT images, our method can successfully segment different abdominal organs with delineations of the boundaries despite the significant variances in organ appearances.

\subsection{Ablation analysis}

\subsubsection{Effectiveness of different components}
\begin{figure}[!tbp]
    \centering
    \includegraphics[width =0.9\linewidth]{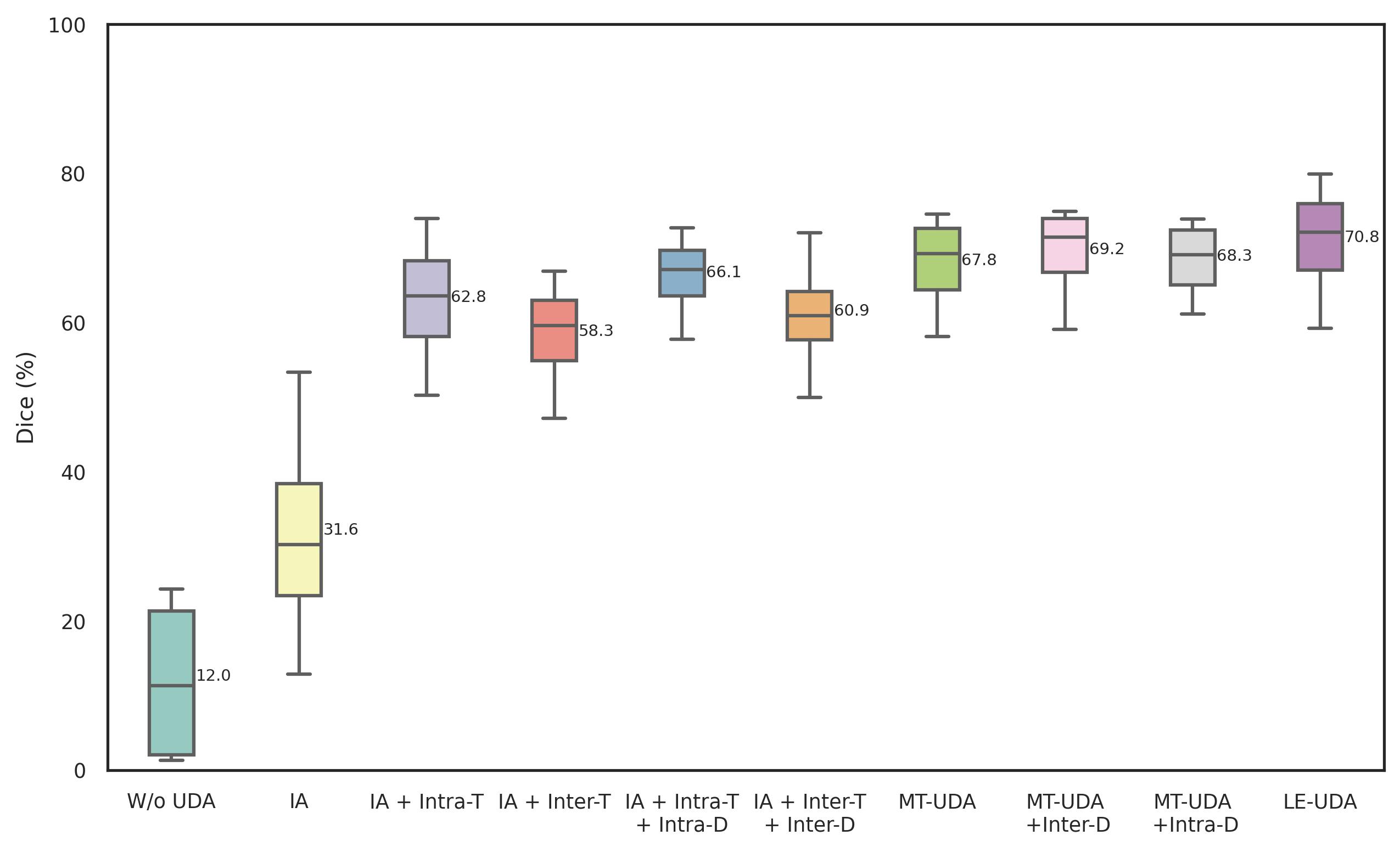}
    \caption{Boxplot of CT cardiac segmentation results produced by different components in~\ourmethod.}
    \label{fig:ablation}
\end{figure}

To evaluate the effectiveness of different components in \ourmethod, we perform ablation studies on CT cardiac segmentation. The ablation studies of key components in \ourmethod~are shown in Fig.~\ref{fig:ablation}. We first evaluate the effectiveness of image alignment via DCAM. More specifically, we directly tested the model trained with limited labeled data $x^{s}$ and $x^{s \rightarrow t \rightarrow s}$,~\emph{i.e.}, W/o UDA, and then tested on synthetic source data $x^{t \rightarrow s}$,~\emph{i.e.}, IA-baseline. Compared to the W/o UDA, image alignment~(IA) can help close the domain gap, but has considerable room for improvement since no feature adaptation or knowledge transfer is explored. We then introduce the source-like data $x^{t \rightarrow s}$ and construct an intra-domain teacher~(Intra-T) for within-domain transfer. Remarkably, the Dice scores are dramatically improved to $62.8\%$, which means that the synthetic data can offer rich information to exploit for addressing both the domain shift and label scarcity issues. Instead of within-domain knowledge transfer, we adopt an inter-domain teacher~(Inter-T) to control the structural consistency between all different pairs from different domains. It is observed that the performance is improved, which is because cross-domain pairs inherit rich domain-invariant structural information, and the information can be transferred to improve the adaptation performance.
We demonstrate the feasibility of involving  SSL in UDA to address both domain shift and source label scarcity. Ultimately, we integrate both within-domain and cross-domain knowledge transfer together,~\emph{i.e.}, MT-UDA, achieving considerable improvement on cross-modality segmentation performance. 

To further validate the effectiveness of the proposed self-ensembling adversarial learning strategy, we first apply different discriminators,~\emph{i.e.,} Inter-D or Intra-D to one teacher alone for feature alignment~(FA). We can see that Inter-D and Intra-D lead to $2.6\%$ and $3.3\%$ enhancements in mean Dice compared to IA + Inter-T and IA + Intra-T, respectively, indicating that either inter-domain feature alignment or intra-domain feature alignment would help close the domain gaps at feature level and guide the knowledge transfer. We further individually implemented Inter-D and Intra-D into the dual-teacher framework, MT-UDA, and a consistent improvement in performance was observed with the inclusion of self-ensembling adversarial learning. The continued increase in segmentation performance shows the effectiveness of adversarial feature alignment. When Inter-D was absent,~\emph{i.e.,} MT-UDA + Intra-D, inter-domain structural knowledge transfer cannot align the domain-invariant discriminative features well, deteriorating the adaptation performance by a $2.5\%$ decrease in mean Dice compared to LE-UDA. With the absence of Intra-D,~\emph{i.e.,} MT-UDA + Inter-D, we observed smaller performance drops due to the smaller domain shifts between the source domain $\mathcal{D}_{s}$ and the source-like domain $\mathcal{D}_{s}^s$. Finally, by incorporating both Inter-D and Intra-D, LE-UDA can effectively improve cross-domain segmentation performance, achieving a 3\% improvement as compared to MT-UDA.

\begin{table*}[t]
\centering
\caption{Ablation Study: Influence of different original and synthetic images in the proposed method for CT cardiac segmentation.}
\setlength\tabcolsep{3pt}
\scalebox{1}{

\begin{tabular}{c|cc|ccc|cccc|cc} 
\toprule
\multirow{2}{*}{Method}               & \multicolumn{2}{c|}{Supervised-only}      & \multicolumn{3}{c|}{Within-domain Transfer}                         & \multicolumn{4}{c|}{Cross-domain Transfer}   & \multirow{2}{*}{Dice} & \multirow{2}{*}{ASD}  \\ 
\cline{2-10}
                     & $x^s$          & $x^{s\rightarrow t \rightarrow s}$                         & $x^s$         & $x^{s\rightarrow t \rightarrow s}$        & $x^{t \rightarrow s}$                 & $[x^s , x^{s\rightarrow t}]$      & $[x^{s\rightarrow t\rightarrow s}, x^{s\rightarrow t}]$   & $[x^{t\rightarrow s}, x^{t}]$      & $[x^{t\rightarrow s}, x^{t \rightarrow s \rightarrow t}]$    &      &       \\ 
\hline
\multirow{2}{*}{No Teacher}           & \checkmark &                                  &            &                                  &                     &                            &                                                     &                           &                                                       & 31.5                  & 17.3                  \\
                                      & \checkmark & \checkmark                       &            &                                  &                     &                            &                                                     &                           &                                                       & 31.6                  & 15.1                  \\ 
\hline
\multirow{2}{*}{Intra-domain Teacher} & \checkmark & \checkmark                       & \checkmark & \checkmark                       &                     &                            &                                                     &                           &                                                       & 47.3                  & 18.2                  \\
                                      & \checkmark & \checkmark                       & \checkmark &                                  & \checkmark          &                            &                                                     &                           &                                                       & 62.8                  & 11.3                  \\ 
                                      & \checkmark & \checkmark                       & \checkmark &                        \checkmark          & \checkmark          &                            &                                                     &                           &                                                       & 57.2                  & 12.9                  \\                             
\hline
\multirow{2}{*}{Inter-domain Teacher} & \checkmark & \checkmark                       &            &                                  &                     & \checkmark                 &                                                     & \checkmark                &                                                       & 55.7                  & 14.1                  \\
                                      & \checkmark & \checkmark                       &            &                                  &                     & \checkmark                 & \checkmark                                          & \checkmark                & \checkmark                                            & 58.3                  & 13.4                  \\ 
\hline
\multirow{3}{*}{\begin{tabular}[c]{@{}c@{}}Dual Teacher\\(MT-UDA)\end{tabular}}         & \checkmark & \checkmark                       & \checkmark & \checkmark                       &                     & \checkmark                 &                                                     & \checkmark                &                                                       & 65.5                  & 9.2                   \\
                                      & \checkmark & \checkmark                       & \checkmark &                                  & \checkmark          & \checkmark                 &                                                     & \checkmark                &                                                       & 66.7                  & 8.0                     \\
                                      & \checkmark & \checkmark                       & \checkmark &                                  & \checkmark          & \checkmark                 & \checkmark                                          & \checkmark                & \checkmark                                            & 67.8                  & 9.5                   \\
                                      & \checkmark & \checkmark                       & \checkmark &                      \checkmark  &           \checkmark          & \checkmark                 & \checkmark                                          & \checkmark                & \checkmark                                            & 66.4                  & 8.6                  \\
\hline
\begin{tabular}[c]{@{}c@{}}Dual Adversarial Teacher\\(LE-UDA)\end{tabular}        & \checkmark & \checkmark                       & \checkmark &                        &           \checkmark          & \checkmark                 & \checkmark                                          & \checkmark                & \checkmark                                            & 70.8                  & 9.6                  \\ 
\bottomrule
\end{tabular}
}
\label{tab:data}
\end{table*}

\subsubsection{Influence of different original and synthetic images}
We further analyze the effects of different original and synthetic images used in our methods. The results of multiple variants of the proposed method with different original and synthetic images are presented in Table~\ref{tab:data}. Going down the rows, we can observe that the baseline performance can be slightly improved by including labeled cycle source data. The marginal improvement could be due to the generated noise within the reconstructed images through the image translation process, which would in turn increase the regularization effect.
With respect to the intra-domain teacher, only using synthetic images from the source domain can also substantially improve the performance from $31.6\%$ to $47.3\%$, since the noise introduced by synthetic images can be also utilized for consistency regularization in the self-ensembling learning process. By introducing synthetic images from the target domain, a significant improvement in performance is observed, which could be attributed to the similarity between the distribution of the synthetic source-like images and the testing images, closing the domain gap between the source domain and the target domain. Interestingly, by including both synthetic cycle source images $x^{s\rightarrow t \rightarrow s}$ and source-like images $x^{t \rightarrow s}$, a drop in performance would be observed. A possible explanation could be the decrease in proportion of source-like images $x^{t \rightarrow s}$, subsequently limiting the effectiveness of consistency regularization and influencing within-domain knowledge transfer for feature adaptation.

With respect to the inter-domain teacher, on the other hand, we can see that all different types of image pairs can contribute to improving the cross-domain adaptation performance. We believe there could be two reasons for this observation. Firstly, self-ensembling learning can effectively leverage unlabeled image pairs for structural knowledge distillation from the inter-domain teacher model to the student model, which is helpful for cross-modality segmentation. Secondly, synthetic images with diverse data distributions contain noise and uncertainty created from the image generation process, thereby enhancing regularization and improving the generalization ability.
However, the inter-domain teacher is still unable to outperform the intra-domain teacher due to the shortage of source labels, since the intra-domain teacher can help close the performance gap by leveraging unlabeled source data. We combine the optimal types of data together into the dual-teacher network, MT-UDA and observe promising performance on unsupervised domain adaptation under source label scarcity. Finally, our LE-UDA further advances dual-domain adversarial learning into self-ensembling teacher-student network, achieving significant performance gains.

\begin{table}[tbp]
\centering
\caption{Sensitivity analysis with respect to hyperparameters $\lambda_{con}^{intra}$, $\lambda_{con}^{inter}$ and $\lambda_{ {adv}}$.}
    \setlength{\tabcolsep}{3mm}{
    \begin{tabular}{cccccc} 
    \toprule
    \multicolumn{6}{c}{\text{Cardiac MRI} → \text{Cardiac CT}}                       \\ 
    \hline
    $\lambda_{con}^{intra}$ & 0.1   & 0.5   & 1    & 5    & 10    \\
    Dice (\%)       &    68.5   & 69.9  & 70.8 & 69.6 & 70.4  \\ 
    \hline\hline
    $\lambda_{con}^{inter}$ & 0.01  & 0.05  & 0.1  & 0.5  & 1     \\
    Dice (\%)        &   69.3    & 70.2  & 70.8 &   68.8   &    67.3   \\ 
    \hline\hline
    $\lambda_{adv}$       & 0.001 & 0.005 & 0.01 & 0.05 & 0.1   \\
    Dice (\%)       &   68.0 &  68.6  & 70.8 & 64.2     & 63.6      \\
    \bottomrule
    \end{tabular}
    }
\label{tab:params}
\end{table}

\subsection{Sensitivity analysis}

\subsubsection{Sensitivity to hyperparameters}
In our experiments, we apply grid search to select the optimal value for $\lambda_{con}^{intra}$, $\lambda_{adv}^{intra}$, $\lambda_{con}^{inter}$ and $\lambda_{adv}^{inter}$. We conduct multiple controlled experiments,~\emph{i.e.,} varying a single hyperparameter while keeping the rest constant each time, to analyze the effects of hyperparameter settings on model performance. For simplicity, we set the same value for $\lambda_{adv}^{inter}$ and $\lambda_{adv}^{intra}$, generally called $\lambda_{adv}$. As observed in Table~\ref{tab:params}, the segmentation performance is relatively stable when the weight of intra-domain consistency loss $\lambda_{con}^{intra}$ is kept between 0.5 and 10, except $\lambda_{con}^{intra} =0.1$ when the performance is relatively lower compared to others. Therefore, we set $\lambda_{con}^{intra}$ to 1 in all experiments. As for the weight of inter-domain consistency loss $\lambda_{con}^{inter}$, the model is not sensitive as the values of $\lambda_{con}^{inter}$ vary between 0.01 to 0.5, and the best performance is achieved when the value is set to 0.01. We further involve adversarial learning and vary the weights of adversarial loss $\lambda_{adv}$ between 0.001 and 0.1. We observe that when the value of $\lambda_{adv}$ is set to 0.005 or 0.01, the performance is relatively stable. However, further increasing or decreasing $\lambda_{adv}$ (\emph{e.g.,} 0.001, 0.05, 0.1) leads to the performance degeneration and unstable adversarial training. We therefore set $\lambda_{adv}$ to 0.01 in all our experiments.

\begin{table}[ht]
\centering

\caption{Sensitivity analysis with respect to layers used in self-ensembling adversarial learning. }
\setlength{\tabcolsep}{2.3mm}{

\begin{tabular}{c|ccccc} 
\hline
\multicolumn{6}{c}{Cardiac MRI →  Cardiac CT}          \\ 
\hline
Layer $i$     & \{1\} & \{1, 2\} & \{1, 3\} & \{1, 4\} & \{1, 5\}  \\ 
\hline
Dice (\%) & 68.0  & 68.5        & 70.8    & 69.8    & 69.6     \\
\hline
\end{tabular}
}
\label{tab:layers}
\end{table}

\subsubsection{Sensitivity to auxiliary layers}
We adopt two layers for multi-level feature consistency and adaptation in our experiments. The performance of connecting different auxiliary layers is shown in Table~\ref{tab:layers}. As highlighted from the table, including auxiliary layers can help improve the segmentation performance by contributing to feature alignment at different levels. Marginal improvements can be observed by making auxiliary predictions from the second last block of U-Net and the best performance is achieved by connecting the third last block of the decoding stage of U-Net to auxiliary classifiers and discriminators. This could be due to the fact that the feature map with moderate scale from the middle layer can help adapt the low-level features, thereby improving cross-domain adaptation performance.

\section{Discussion}
\begin{figure}[tbp]
    \centering
    \includegraphics[width =0.9\linewidth]{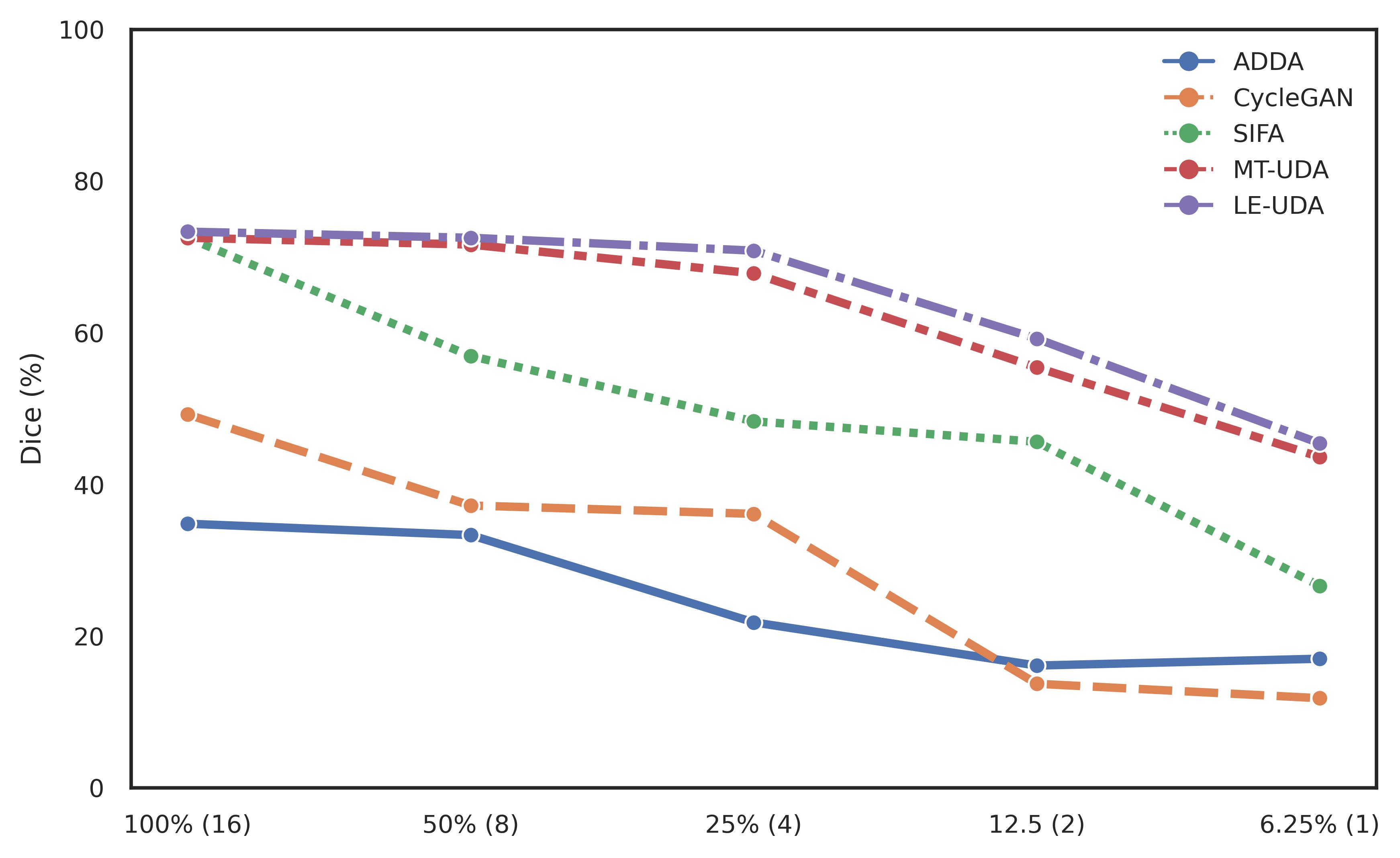}
    \caption{CT cardiac segmentation results of various approaches under different source label ratios (number of labels).}
    \label{fig:varshot}
\end{figure}

\begin{figure}[tbp]
    \centering
    \includegraphics[width =\linewidth]{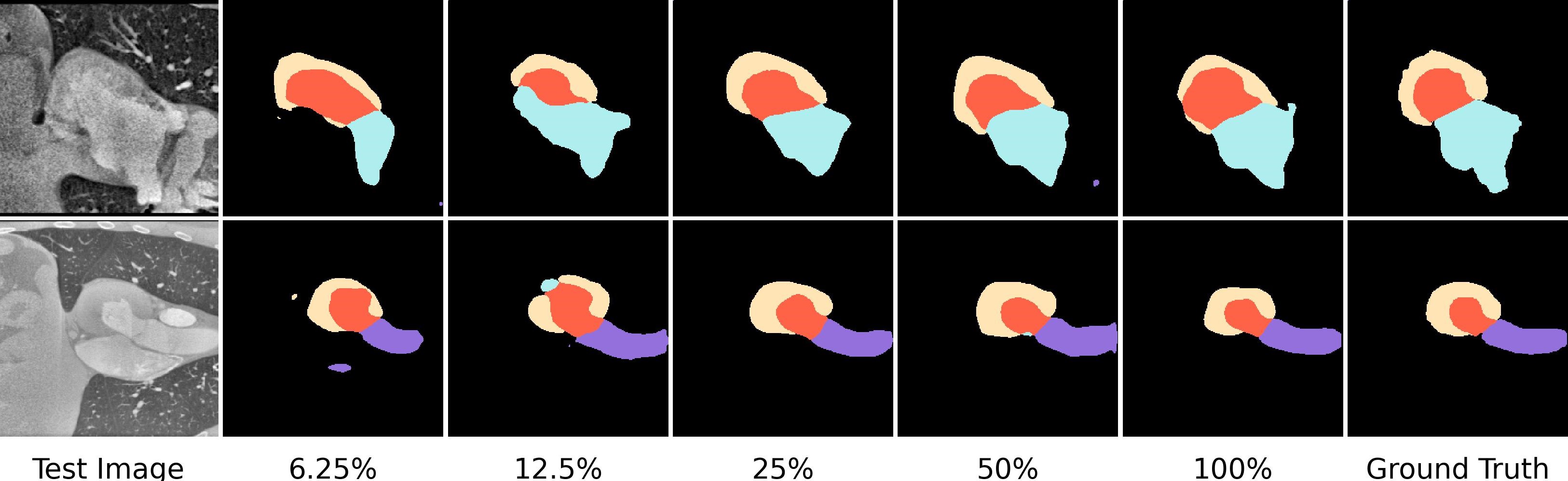}
    \caption{Visual comparison of segmentation results of LE-UDA under different source label ratios for
cardiac CT images with four substructures,~\emph{i.e.,} AA~(purple), LAC~(blue), LVC~(red) and MYO~(yellow).
}
    \label{fig:allratio}
\end{figure}

Domain shift has been a prevalent issue in deep learning-based cross-domain medical image segmentation. Various unsupervised domain adaptation~(UDA) methods~\cite{zhu2017unpaired,russo2018source,zhang2018task, bousmalis2017unsupervised, zhao2018supervised, ganin2016domain, tzeng2017adversarial,hoffman2018cycada,chen2019synergistic} have been proposed to narrow the domain gap from different perspectives, achieving significant cross-domain adaptation performances in medical image segmentation, but neglecting the existence of source label scarcity in the real-world scenarios. In this work, we identify a novel UDA problem hidden in medical image segmentation, which is UDA under source label scarcity and propose a novel method named ``Label-Efficient Unsupervised Domain Adaptation"~(LE-UDA) to address the aforementioned issue. Based on our previous work~\cite{zhao2021mt}, we further propose to integrate adversarial feature learning into self-ensembling teacher-student network for better feature alignment and reliable knowledge transfer. We showcased the effectiveness of LE-UDA for bidirectional UDA using 25\% source labels on two datasets, receiving smaller performance drops than MT-UDA and other UDA approaches. We further implemented different methods under various source label ratios for a more comprehensive comparison. As demonstrated in Fig.~\ref{fig:varshot}, we can observe that with a decrease in source label ratio, image adaptation~\emph{i.e.}, CycleGAN suffers a higher performance drop than feature adaptation,~\emph{i.e.}, ADDA, which can be attributed to model overfitting on the single fake sample without any other regulations, while feature adaptation induces adversarial learning on the feature space, diversifying the feature distribution against domain shift. Nevertheless, SIFA receives significant performance reduction when training with only one source label. In comparison, our method can consistently achieve better segmentation results under different source label ratios, even on the extremely low labeled data regime,~\emph{i.e.}, one-shot UDA. Fig.~\ref{fig:allratio} highlights the quantitative comparison by LE-UDA using different source label ratios. As illustrated in the figure, the segmentation results by LE-UDA can mostly maintain the shapes of different substructures, which further demonstrates the feasibility of LE-UDA against source label scarcity for UDA.

\begin{table}[ht]
\centering
\setlength\tabcolsep{4pt}
\caption{Complexity and performance of different segmentation backbones for cardiac substructure segmentation.}
\scalebox{0.85}{
\begin{tabular}{c|c|ccccc} 
\toprule
\multicolumn{7}{c}{Cardiac CT}                                                                                            \\ 
\hline
\multirow{2}{*}{Method} & \multirow{2}{*}{\# Params (M)} & \multicolumn{5}{c}{Dice (\%)}                                  \\ 
\cline{3-7}
                        &                                     & AA         & LAC       & LVC       & MYO       & Average  \\ 
\hline
SIFA                    &         27.53                         & 93.5(3.3)  & 91.8(1.9) & 90.4(2.9) & 83.7(8.1) & 89.8     \\
Ours                    &       28.95                              & 85.5(13.2) & 88.6(3.4) & 88.1(4.9) & 84.2(5.6) & 86.6     \\ 
\bottomrule
\multicolumn{1}{c}{}    & \multicolumn{1}{l}{}                &            &           &           &           &          \\ 
\toprule
\multicolumn{7}{c}{Cardiac MRI}                                                                                           \\ 
\hline
\multirow{2}{*}{Method} & \multirow{2}{*}{\# Params (M)}  & \multicolumn{5}{c}{Dice (\%)}                                  \\ 
\cline{3-7}
                        &                                     & AA         & LAC       & LVC       & MYO       & Average  \\ 
\hline
SIFA                    &       27.53                            & 81.2(5.6)  & 83.5(9.2) & 93.1(2.0) & 80.0(4.5) & 84.5     \\
Ours                    &       28.95                              & 84.2(3.4)  & 82.8(8.6) & 92.3(1.7) & 80.4(3.8) & 84.9     \\
\bottomrule
\end{tabular}
}
\label{tab:backbone}
\end{table}
\begin{figure}[tbp]
    \centering
    \includegraphics[width =0.8\linewidth]{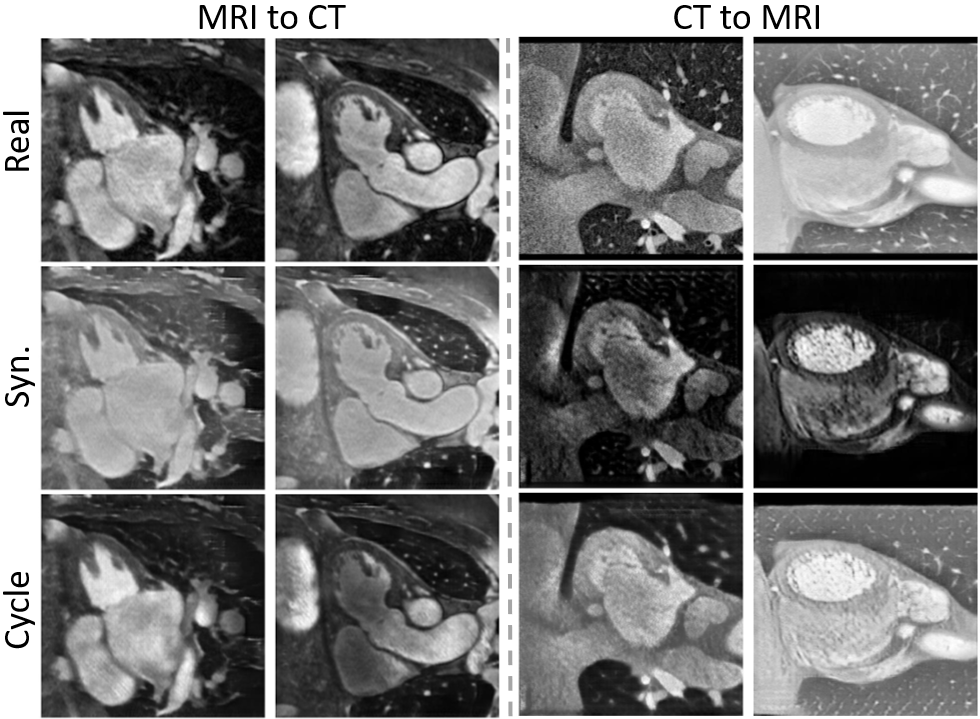}
    \caption{Example outputs of our translation from MRI to CT (first two columns) and from CT to MRI (last two columns).
}
    \label{fig:gan_image}
\end{figure}

In our proposed framework, different types of images are synthetically generated by the dual cycle alignment module (\emph{i.e.,} CycleGAN~\cite{zhu2017unpaired}), which not only improve the diversity of the training data but also boost the generalization capability of the model. Fig.~\ref{fig:gan_image} shows some typical synthetic (syn.) and reconstructed (cycle) images generated by CycleGAN. As can be observed visually, most generated images are close to real images, maintaining similar structures and shapes. However, noises may be introduced into the training process in despite of enforcing the preservation of semantics and contents between the transformed images and the original images via cycle-consistency loss. Some transformed images of CycleGAN (\emph{e.g.,} 4th column in Fig.~\ref{fig:gan_image}) have artifacts, checkerboard effects, and missing anatomies (\emph{e.g.,} ascending aorta). In LE-UDA, different perturbations such as noises are introduced to the student and teacher networks to encourage their predictions to be consistent, hence the influence of noises induced by CycleGAN could be decreased to some extent. One possible limitation is that image pairs with inconsistent anatomical structures may influence the inter-domain structural knowledge transfer. Addressing the uncertainty of GAN-based medical image translation is still an open problem. Many efforts~\cite{zhang2018translating, zeng2020icmsc, shin2022cosmos} have been devoted to addressing the limitations of GANs on uncertainty and consistency. We believe that improved performance could be obtained if we were to more effectively map the pixels between the generated image and the original image. To further improve the performance, we will explore this issue in our future work. For computational efficiency, we adopt the widely used segmentation network, 2D U-Net for our experiments. To evaluate the effectiveness of our segmentation backbone, we compare the performance of U-Net with the segmentation backbone of SIFA~\cite{chen2020unsupervised} for cardiac segmentation, as reported in Table~\ref{tab:backbone}. It is observed that U-Net can achieve comparable performance with similar number of parameters to the SIFA backbone. Compared to SIFA, which requires a shared encoder for both segmentation and translation, our framework is generic and more flexible of modifying segmentation network architectures.  Moreover, it may be beneficial to adopt 3D networks for medical image synthesis and segmentation. However, a 3D implementation of our framework consisting of multiple components would be highly memory intensive. As such, we plan to tackle this issue in our future work.

\section{Conclusions}
In this paper, we propose a novel unsupervised domain adaptation framework,~\ourmethod, which addresses both domain shift and source label scarcity for cross-domain medical image segmentation.  In~\ourmethod, two teacher models are employed to explore both inter-domain knowledge and intra-domain knowledge hidden behind multiple data sources with diverse distributions for boosting cross-domain segmentation performance. By integrating adversarial learning into the dual-teacher network, the performance is further improved. The experimental results and ablation analysis reveal that ~\ourmethod~is effective, outperforming other UDA methods on unsupervised domain adaptation against source label scarcity by a significant margin.

\bibliographystyle{IEEEtran}
\bibliography{references.bib}

\end{document}